\documentclass[fleqn,usenatbib]{mnras}

\usepackage{newtxtext,newtxmath}
\usepackage[T1]{fontenc}
\usepackage{ae,aecompl}
\usepackage{pdflscape}

\usepackage{graphicx}	% Including figure files
\usepackage{amsmath}	% Advanced maths commands

\usepackage{amssymb}	% Extra maths symbols
\usepackage{xcolor}
\usepackage{hyperref}
\usepackage{orcidlink}

\newcommand{\orcid}[1]{\href{https://orcid.org/#1}{\textcolor[HTML]{A6CE39}{\aiOrcid}}}

\def\la{\mathrel{\mathpalette\fun <}}
\def\ga{\mathrel{\mathpalette\fun >}}
\def\fun#1#2{\lower3.6pt\vbox{\baselineskip0pt\lineskip.9pt
        \ialign{$\mathsurround=0pt#1\hfill##\hfil$\crcr#2\crcr\sim\crcr}}}

\title[BOSS galaxy clustering] {Small scale clustering of BOSS galaxies: dependence on luminosity, color, age, stellar mass, specific star formation rate and other properties}

\author[Z. Zhai et al.]{\parbox{\textwidth}{
Zhongxu Zhai,$^{1,2,3,4}$\thanks{E-mail: zhongxuzhai@sjtu.edu.cn}
Will J. Percival, $^{3,4,5}$
Hong Guo, $^{6}$
}\vspace*{4pt}\\
$^{1}$Department of Astronomy, School of Physics and Astronomy, Shanghai Jiao Tong University, Shanghai 200240, China \\
$^{2}$Shanghai Key Laboratory for Particle Physics and Cosmology, Shanghai 200240, China \\
$^{3}$Waterloo Center for Astrophysics, University of Waterloo, Waterloo, ON N2L 3G1, Canada \\
$^{4}$Department of Physics and Astronomy, University of Waterloo, Waterloo, ON N2L 3G1, Canada \\
$^{5}$Institute of Cosmology and Gravitation, University of Portsmouth, Burnaby Road, Portsmouth, PO1 3FX, United Kingdom\\
$^{6}$Key Laboratory for Research in Galaxies and Cosmology, Shanghai Astronomical Observatory, Shanghai 200030, China \\ 
}
\date{Accepted XXX. Received YYY; in original form ZZZ}

\pubyear{2020}

\begin{document}
\label{firstpage}
\pagerange{\pageref{firstpage}--\pageref{lastpage}}
\maketitle

\begin{abstract}

We measure and analyze galaxy clustering and the dependence on luminosity, color, age, stellar mass and specific star formation rate using Baryon Oscillation Spectroscopic Survey (BOSS) galaxies at $0.48<z<0.62$. We fit the monopole and quadrupole moments of the two-point correlation function (2PCF) and its projection on scales of $0.1$ -- $60.2h^{-1}$Mpc, after having split the catalog in a variety of ways. We find that the clustering dependence is consistent with previous well-established results showing the broad trends expected: For example, that brighter, redder, older, more massive and quenched galaxies are more strongly clustered. We also investigate the dependence on additional parameters previously derived from stellar population synthesis model fits to the spectra. We find that galaxy clustering depends on look-back formation time at a low level, while it has little dependence on metallicity. To understand the physics behind these trends, we fit the clustering with a simulation-based emulator to simultaneously model cosmology and galaxy bias using a Halo Occupation Distribution framework. After marginalizing parameters determining the background cosmology, galaxy bias, and a scaling parameter to decouple halo velocity field, we find that the growth rate of large scale structure as determined by the redshift-space distortions is consistent with previous analysis using the full sample, and we do not find evidence that cosmological constraints depend systematically on galaxy selection. This demonstrates that cosmological inference using small scale clustering measurements is robust to changes in the catalog selection. 

\end{abstract}


\section{Introduction}

Large galaxy surveys from the Sloan Digital Sky Survey (SDSS) provide an incredible resource for the observational study of cosmology. Data from SDSS-I/II, \citep{SDSS_York, Abazajian_2009}, the  Baryon Oscillation Spectroscopic Survey (BOSS; \citealt{Dawson_BOSS}), and extended BOSS (eBOSS; \citealt{eBOSS_Dawson}), have mapped the Universe using millions of galaxies over a wide redshift range enabling accurate measurements of the expansion history of the universe and the growth history of the large scale structure \citep{eboss-cosmo}. In addition, the clustering measurements on small scales strongly constrain the physics of galaxy formation and evolution (\citealt{Zheng_2007, White_2007, Brown_2008}). 

Due to the non-linear evolution of dark matter dynamics, and the coupling with baryonic processes of galaxy formation, it has been challenging to extract cosmological information at small scale. One solution is to construct a flexible emulator based on high resolution N-body simulations coupled with a halo-based galaxy distribution model, which provides predictions of galaxy clustering with variations in both the background cosmology and halo occupation distribution (HOD). In \citet{Zhai_2019, Zhai_2022}, we developed and applied the emulator approach to model the two-point statistics of the BOSS massive galaxies and measured the linear growth rate over a wide redshift range. The emulator is constructed using Gaussian Processes (GP) to model and interpolate galaxy statistics in the high dimensional parameter space. This provides a machinery to allow fast predictions of galaxy statistics for an arbitrary parameter set. When fitting to observational measurements, a dense sampling can quickly be made in the parameter space to provide a well sampled posterior distribution of the model parameters. Similar methods have been used to model the galaxy 2-point correlation function and beyond, including the halo mass function (\citealt{McClintock_2018}), halo bias function (\citealt{McClintock_2019}), non-linear matter power spectrum (\citealt{Heitmann_2010, Knabenhans_2019, Arico_2020}), dark matter halo statistics (\citealt{Nishimichi_2019, Kobayashi_2020, Miyatake_2021}), Lyman$-\alpha$ forest flux power spectrum (\citealt{Simeon_2019, Rogers_2019, Walther_2020, Pedersen_2020}), and galaxy lensing (\citealt{Wibking_2017, Wibking_2020, Salcedo_2020}).

For cosmological measurements from galaxy clustering on small scales, understanding the non-linear correlations between cosmological parameters and galaxy formation is critical to achieve unbiased cosmological measurements \citep{Ross_2014, Patej_2016, Marin_2016, Mohammad_2018}. The emulator approach allows such investigations by simultaneously modeling the dependence of galaxy clustering on both cosmology and galaxy bias. After marginalizing over the cosmological parameters, galaxy bias parameters and a scaling parameter for dark matter halo velocity field, \citet{Zhai_2022} (hereafter Z22) found that the linear growth rate measurement $f\sigma_{8}$ is lower than the prediction from Planck observations (\citealt{Planck_2020}), especially at $z=0.55$ with a tension of higher than $3\sigma$. A similar analysis using eBOSS LRGs at $z=0.7$ with a lower number density also reported a lower measurement of growth rate \citep{Chapman_2021}. Using similar BOSS-CMASS galaxies but an independent emulator model, \cite{Yuan_2022} found that the clustering amplitude is also lower than Planck. A natural question is whether these cosmological measurements are affected by the galaxy selection or galaxy properties. Our goal in the current work is to use this methodology to explore the following questions: (1) is the emulator approach robust, measuring consistent linear growth rates of large scale structure using different sample selections? (2) after marginalizing over cosmological parameters, do the fits to the clustering of populations split in galaxy properties provide insight into the physics of galaxy formation and evolution? 

The dependence of clustering on galaxy properties and its interpretation in terms of galaxy formation and evolution has been the subject of significant interest over a long period of time. Using SDSS galaxies, \cite{Li_2006a} studied how the clustering depends on galaxy properties and found that galaxies with redder colors, larger 4000-\AA\ break strength, higher concentration and larger surface mass densities cluster more strongly. \cite{Swanson_2008} used a counts-in-cells analysis to measure the relative bias of galaxy samples with different luminosities and colors. This was extended to study scale dependent galaxy bias in \cite{Cresswell_2009}. \cite{Zehavi_2011} analyzed the Main Galaxy Sample from SDSS DR7 and measured the projected correlation function to explore the luminosity and color dependence of the clustering measurement, and then interpreted the result in the halo occupation distribution (HOD) modeling assuming a $\Lambda$CDM cosmology. A similar analysis of color and luminosity-dependence at higher redshift was performed in \cite{Guo_2014} using early BOSS-CMASS galaxies.

More recent work using SDSS galaxies can be found in \cite{Shi_2016, Xu_2016, Hearin_2017} and references therein. In addition to luminosity and color, \cite{MonteroDorta2017} found that the clustering of galaxies is also affected by their assembly history, i.e. fast-growing luminous red galaxies (LRGs) are more strongly clustered and reside in denser environments than slow-growing systems. Besides the analyses based on SDSS, \cite{Skibba_2014} utilized the data from PRIMUS sample to measure the luminosity and color dependence of galaxy clustering and found consistent result with earlier measurements. Using the DEEP2 galaxy redshift survey at $z\sim1$, \cite{Coil_2008} measured the  color and luminosity dependence of galaxy clustering, and produced a result roughly consistent with the analysis based on SDSS galaxies. \cite{Mostek_2013} and \cite{Coil_2017} further analyzed the DEEP2 galaxy sample and found a positive correlation between stellar mass and clustering amplitude. They also report that the clustering dependence on star formation rate (SFR) or specific SFR (sSFR) is different for red and blue galaxies. This result is also compared with local SDSS galaxies at $z\sim0.03$ in \cite{Berti_2021}. The investigation of clustering dependence was also performed for high redshift galaxies: \cite{Durkalec_2018} measured the clustering signals using VIMOS Ultra Deep Survey (VUDS) in the redshift range $2<z<3.5$ and found a dependence on stellar mass and luminosity similar to the low redshift galaxies.  

The analyses of these works have relied heavily on the halo model framework \citep{Jing_1998, Peacock_2000, Cooray_2002, Berlind_2002}. In this framework, \cite{Skibba_2009} incorporated a description for the luminosity and color dependence of galaxy clustering and enables investigation of the correlation between halo mass, environment and galaxy color. \cite{Masaki_2013} and \cite{Saito_2016} extended the subhalo abundance matching method to incorporate the assignment of galaxy color to subhalos and investigate its impact on clustering measurement. 

The previous analyses started from the premise that the background cosmology is known. We extend this work using the  emulator based approach to allow us to correctly marginalise over the cosmological model when interpreting HOD-based measurements of small-scale galaxy properties. We also extend the investigations by varying cosmological parameters and splitting the galaxies by intrinsic galaxy properties beyond luminosity and color. We base our analysis on fitting to the BOSS data, making use of the galaxy properties provided in the value-added catalogs (VAC). In addition to the broad band luminosities, these provide galaxy parameters based on the fitting of stellar population (SP) synthesis algorithm and outputs galaxy age, stellar mass, K+E corrected magnitude, sSFR as well as other parameters. 

Our paper is organized as follows. In Section~\ref{sec:data} we introduce the data and galaxy selection algorithm. In Section~\ref{sec:measurement} we describe the measurement of galaxy clustering and modeling. Section~\ref{sec:result} gives the constraint results based on galaxy luminosity, color, age, stellar mass, specific Star Formation Rate and SP parameters. Section~\ref{sec:discussion} discusses our result and Section~\ref{sec:conclusion} presents the conclusion.

\section{Observational data}\label{sec:data}

In this work, we use the same LSS catalog from BOSS (\citealt{Reid_2016}) that we used in our earlier paper Z22, but focus on the high redshift subsample only ($0.48<z<0.62$) as this provided the $f\sigma_{8}$ measurement with the highest tension from the Planck expectation in Z22, compared with samples at lower redshifts. The galaxy properties we use come from the value added catalog (VAC) \citep{Ahn_2014} containing the ``Granada" estimates based on fitting to the SDSS photometry and spectroscopy (\citealt{Conroy_2009, MonteroDorta_2016}) \footnote{The choice of ``Granada" VAC is not unique for the analysis in this work, and we expect that using the ``Portsmouth" or ``Wisconsin" VACs would produce similar results. We chose the ``Granada" VAC simply because it provides a comprehensive set of parameters for individual galaxies that enable a thorough investigation.}. This VAC provides multiple sets of galaxy properties for a variety of modeling choices, including different priors on the star formation model (early formation-time and a wide range of formation times), with dust or without dust, \cite{Salpeter_1955} or \cite{Kroupa_2001} initial mass functions. There are 8 versions of the catalog in total.

For each galaxy property we will consider, we split the sample into thin redshift slices of width $\Delta z=0.005$. Then we rank order galaxies in each thin slice by the relevant galaxy property. We will first investigate the luminosity dependence of our analysis. In this case, the galaxy property in the rank-ordering is i-band magnitude, to be consistent with Z22. We apply three thresholds of number density: 1, 2 and 3$\times10^{-4}[h^{-1}\mathrm{Mpc}]^{-3}$, keeping the brightest galaxies until we reach the required number density threshold (we call these thresholds N1, N2 \& N3). Note that the number density of $2\times10^{-4}[h^{-1}\mathrm{Mpc}]^{-3}$ is equivalent to that used in Z22. 

Following this, we will study the clustering dependence on galaxy properties from the Granada VAC. We match the BOSS LSS catalog with the Granada VAC by ``MJD", ``PLATE" and ``FIBERID", which can uniquely define an object. In this process, we find that <1.5\% of the galaxies in the NGC do not have information in the Granada VAC, while the SGC has a missing fraction of <1.0\%. However the missing galaxies in the SGC are more concentrated on a few SDSS plates, which may induce additional problems for the clustering measurements. The missing galaxies in the NGC are more randomly distributed across the footprint, and the clustering is only weakly affected. Therefore we only use NGC throughout our analysis, including the luminosity dependence measurements. From the Granada VAC, we choose the following galaxy properties in the analysis:
\begin{itemize}
    \item color: R-I from the K+E corrected magnitude. Note that this color definition is also used by \cite{Ross_2014} for color-dependent measurement of distance scale and growth rate, which is similar to our work but focuses on larger, linear scales.
    \item age: Mean mass-weighted average age of the stellar population.
    \item stellar mass: Median stellar mass of galaxy from the SP fit.
    \item sSFR: Mean Specific Star Formation Rate.
    \item LBF: best-fit value of the look-back formation time, one of the SP parameters.
    \item metallicity: best-fit value of metallicity, one of the SP parameters.
\end{itemize}

For each of these properties (i.e. all properties except luminosity), we select galaxies in ascending (descending) order from the lowest (highest) value until each thin slice reaches the threshold of galaxy number density for N1 and N2 thresholds only. This means that in each thin redshift slice, we select the rank-ordered galaxies from two ends, until the number density reaches N1 or N2. Catalogues created in this way up to the N3 threshold would be too similar for a comparative analysis to be useful. The resultant sub-catalogs thus have approximately constant number density across the entire redshift range of interest. Although BOSS is a colour selected sample, many of the properties of galaxies in the population that we consider do not change significantly with redshift. Thus, in many ways they behave as volume limited samples. For example, Fig.~\ref{fig:Bestfit_CF_luminosity} shows that a density cut made as described above for galaxies ordered by absolute luminosity gives a relatively sharp cut as would be expected for a volume limited sample. Galaxy selections from the least massive or youngest end may have properties that weakly depend on redshift, such that the sample is not volume-limited. An investigation with models that can take this weak incompleteness into account could further improve the robustness of our analysis, but we leave this for future work.

\begin{figure}
\begin{center}
\includegraphics[width=9cm]{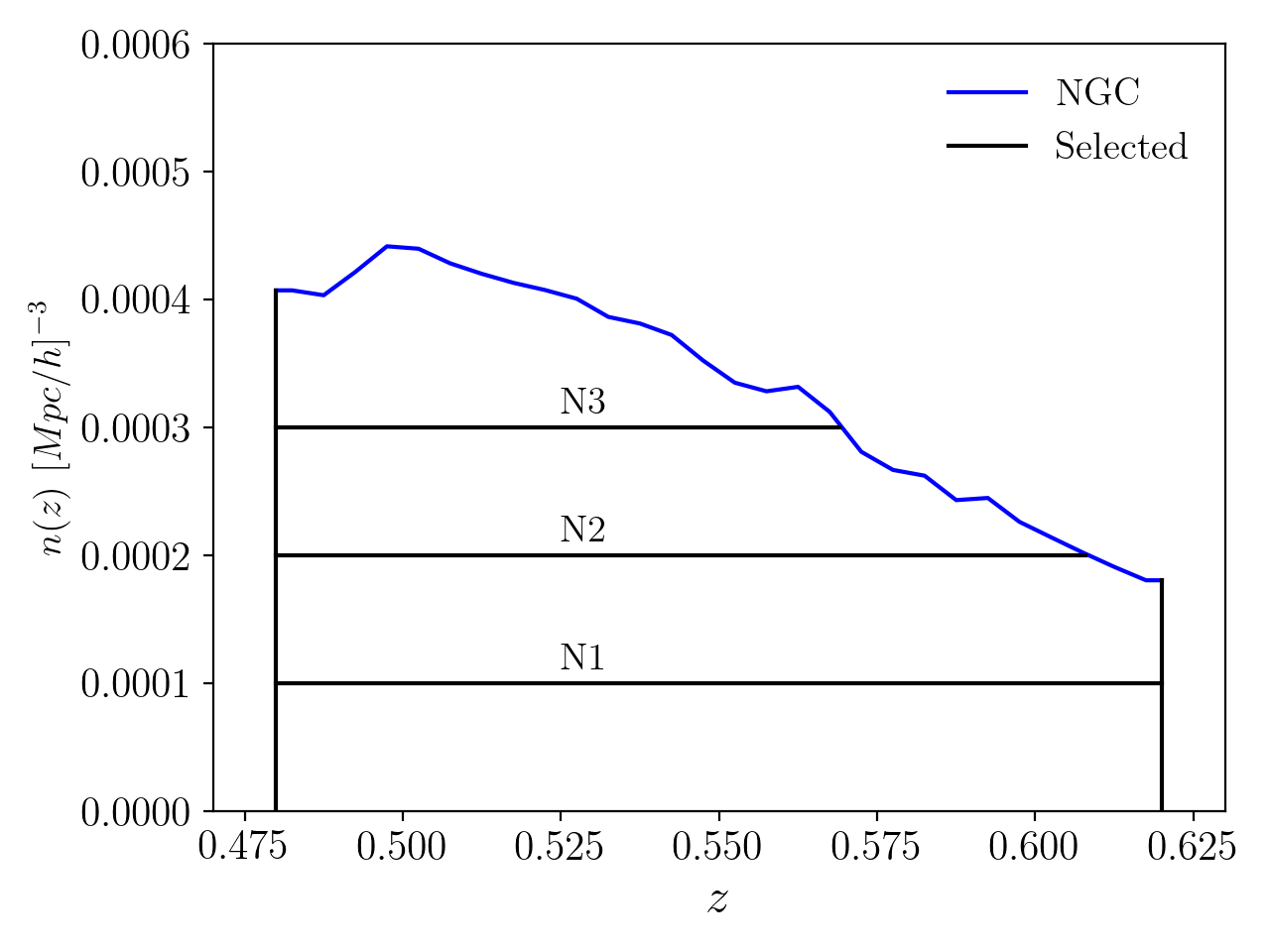}
\caption{The comoving number density of the BOSS DR12 NGC galaxies as a function of redshift. The three horizontal lines mark the threshold of number density in our analysis, as well as the number densities corresponding to the emulator.}
\label{fig:nz}
\end{center}
\end{figure}

In Figure~\ref{fig:nz}, we show the number density of the BOSS NGC sample. The horizontal lines mark 3 thresholds of number density. Note that for threshold of $2\times10^{-4}[h^{-1}\mathrm{Mpc}]^{-3}$, the two extreme samples can have slight overlap which we believe won't significantly affect the qualitative conclusions made from our results. Here the extreme samples refer to the fact that we select galaxies from the two ends, i.e. in the extremal ends for properties except luminosity. For instance, for color, we select from the red end and blue end, we keep adding less red and less blue galaxies until the number density is reached. If the total number of galaxies in a thin redshift slice is less than twice the required number density, some galaxies in the middle (both red and blue) can be selected in both sub-catalogs.

\section{Measurement and Modeling of small scale clustering}\label{sec:measurement}

We use 2-point correlation function to measure galaxy clustering in this analysis. The methodology follows Z22 so we only summarise it here.

\subsection{Measurement from BOSS galaxies}

We use the estimator from \citet{LS_1993} to compute the correlation function for each galaxy sample considered,
\begin{equation}
    \xi(\vec{r}) = \frac{DD-2DR+RR}{RR},
\end{equation}
where DD, DR and RR are numbers of data-data, data-random and random-random pairs of galaxies with a separation vector $\vec{r}$. We use the Planck 2016 $\Lambda$CDM cosmology to convert redshift to Cartesian coordinates with $\Omega_{m}=0.307$. We correct the ``Fiber Collision" effect with the method by \cite{Guo_2012} to account for the missing galaxies in the catalog due to loss of fiber assignment. This correction method has been demonstrated to recover the correlation function on scales below and above the collision scale to reasonable accuracy. Alternative method to correct this fiber collision can be found in \cite{Hahn_2017, Percival_2017} and \cite{Bianchi_2017}, and has been used in the clustering measurements of eBOSS data \citep{Mohammad_2020, Chapman_2021}.

We compress the correlation function into the projected correlation function and redshift space multipoles. For the former, we include it to mitigate the redshift space distortions and evaluate the real space galaxy bias
\begin{equation}
    w_{p}(r_{p}) = 2\int_{0}^{\infty}d\pi \xi(r_{p}, \pi),
\end{equation}
where the correlation function has been decomposed on a 2D grid of separations perpendicular ($r_{p}$) and parallel ($\pi$) to the line of sight. The redshift space multipoles are obtained through Legendre decomposition
\begin{equation}
    \xi_{\ell} = \frac{2\ell+1}{2}\int_{-1}^{1}L_{\ell}(\mu) \xi(s, \mu) d\mu,
\end{equation}
where $L_{\ell}$ is the Legendre polynomial of order $\ell$, and the coordinates of the correlation function are related through
\begin{equation}
    s=\sqrt{\pi^{2}+r_{p}^{2}}, \quad \mu=r_{p}/s.
\end{equation}
In our analysis, the measurement of galaxy correlation function is performed with 10 logarithmic bins for $r_{p}$ or $s$ from 0.1 to 60.2 $h^{-1}$Mpc, and thus each statistic $w_{p}$ or $\xi_{\ell}$ has 9 data points.

\subsection{Modeling}

For the theoretical modeling of the galaxy correlation function, we use the HOD model presented in Z22 to statistically describe the galaxy population within dark matter halos. This HOD model has basic parameters for the mass dependence of central and satellite occupancies respectively, see e.g. \cite{Zheng_2005, CMASS_Martin, Parejko_LOWZ, Zhai_2017} and references therein. Then we have one parameter to control the concentration of the satellite distribution relative to the NFW profile (\citealt{NFW_1996}), two parameters for the velocity bias of centrals and satellites, one parameter $\gamma_{f}$ to scale and decouple the velocity field of dark matter halos predicted by the $w$CDM+GR simulations, and three parameters to model the assembly bias caused by local environment. {We note that the modeling of assembly bias is not unique. 
%It is possible to implement this effect using properties such as halo age, concentration and so on. 
However, recent studies using numerical simulations show that the local environment of the halo at the present day is an excellent indicator of assembly bias, see e.g. \cite{Han_2019, Yuan_2020,Xu_2020} and references therein. Therefore we apply a model in the current analysis, following Z22, with an assembly bias parameter that scales the HOD parameter $M_{\text{min}}$ via
\begin{equation}\label{eq:Mmin_AB}
    \bar M_{\rm min} = M_{\text{min}}\left[1 + f_{\rm env}\text{erf}\left(\frac{\delta-\delta_{\rm env}}{\sigma_{\rm env}}\right)\right],
\end{equation}
where $f_{\text{env}}, \delta_{\text{env}}$ and $\sigma_{\text{env}}$ control the overall strength of halo environment on the galaxy distribution. We also include the parameter $f_{\text{max}}$ throughout our analysis, which describes the asymptotic behavior of the central occupancy, i.e. the most massive halos can have less than than 1 galaxy at the center on average. This extra degree of freedom is important when we subsample the galaxies, leaving only a fraction of the haloes populated by a central. Although this parameter adds more degrees of freedom compared with the original model from \cite{Zheng_2005}, we note that the asymptotic behavior of central occupancy at the high mass end, may not be ideal for some of the subsamples such as less massive or younger galaxies. Their central occupancy may drop quickly for very massive halos. However, our results in later sections seem to show that the current HOD model doesn't introduce significant bias for cosmological measurements. The above description fully defines our HOD model with great flexibility and we term the parameters as galaxy bias parameters in the following analysis.

Using the HOD model to connect dark matter halos and galaxies, we build the emulator with the Aemulus suite (\citealt{DeRose_2018}) of simulations. The Aemulus suite has 40 simulations with different cosmologies, which we use to train the emulator and another 35 simulations that we use to test the performance. All the simulations have a box size of $1.05h^{-1}$Gpc with $1400^3$ dark matter particles. The particle mass is $\sim3.5\times10^{10}h^{-1}\text{M}_{\odot}$ with a slight dependence on cosmology and this makes the resolution suitable for massive galaxy populations like the BOSS-CMASS sample. The dark matter halos are identified by the ROCKSTAR spherical overdensity halo finder (\citealt{Behroozi_2013b}) and we choose M$_{200b}$ as the mass definition for our analysis. The resultant emulator is an algorithm to make fast and accurate predictions for $w_{p}, \xi_{0}$ and $\xi_{2}$ in the high dimensional parameter space for $w$CDM+GR+galaxy bias. The $w$CDM model is a minimal extension of $\Lambda$CDM model with the equation of state parameter of dark energy $w$ as a free parameter. The other cosmological parameters of our model include the matter density $\Omega_{m}$, the baryon energy density $\Omega_{b}$, the amplitude of matter fluctuations $\sigma_{8}$, the dimensionless Hubble parameter $h$, the spectral index of the primordial power spectrum $n_{s}$ and the number of relativistic species $N_{\text{eff}}$. In addition, we do not correct for the offset in the distance-redshift relationship between model being fitted and fiducial model (the Alcock--Paczynski  effect, \citealt{Alcock_1979}), matching the methodology of \citet{Zhai_2022}. This only has a minor effect on the results at small scales due to the lack of features in the clustering measurement (e.g. the results in table~2 of \citealt{Chapman_2021}).

In order to investigate the impact from various galaxy properties, we build emulators with three number densities N1, N2 and N3 corresponding to 1, 2 and 3$\times10^{-4}[h^{-1}\mathrm{Mpc}]^{-3}$ as introduced in Section~\ref{sec:data}. The clustering analysis is performed with matched number density for the measurement from BOSS galaxies and emulator.

\subsection{likelihood analysis}

We obtain the posterior of the unknown parameters assuming a Gaussian likelihood, which is defined as 
\begin{equation}
\ln{\mathcal{L}} = -\frac{1}{2}(\xi_{\text{emu}}-\xi_{\text{obs}})C^{-1}(\xi_{\text{emu}}-\xi_{\text{obs}}),
\end{equation}
where $C$ is the covariance matrix, $\xi_{\text{emu}}$ and $\xi_{\text{obs}}$ are galaxy correlation functions from emulator prediction and BOSS galaxies respectively. We use the nested sampling algorithm  (\citealt{Skilling_2004}) provided by \textmd{MultiNest} (\citealt{Feroz_2009, Buchner_2014}) package to explore the likelihood surface. The output can give the Bayes Evidence for model selection, as well as the posterior distribution of parameters of interest. Following Z22, our Multinest run has 1000 live-points throughout the calculation. Most of the other parameters are set as default, including sampling efficiency=0.8 and the constant efficiency mode is turned off. The analysis is stopped when it yields converged results for the evidence. For a typical analysis in this work, it needs roughly five thousand CPU hours to finish. A thorough exploration of the sampling algorithm can be found in \cite{Lemos_2023}. The sampling also needs prior information for model parameters. We employ the same method as Z22 to define a restricted area for the cosmological parameters based on the training cosmologies from the Aemulus suite, and flat priors for galaxy bias parameters. 

Due to the inaccuracies of the emulator predictions, the covariance matrix entering the analysis has two contributions, sample variance and emulator error.
\begin{equation}
C = C_{\text{sam}}+C_{\text{emu}}.
\end{equation}
In this work, we adopt the method from \citet{Lange_2021} to reduce the effect of noise by smoothing the covariance matrix. To start, we measure the covariance matrix of galaxy correlation function by jackknife-resampling. Then we normalize it by the diagonal element to get the correlation matrix, and replace the diagonal elements by the average of the neighbouring elements, and smooth the off-diagonal elements with a two-dimensional Gaussian kernel. Next we replace back the diagonal elements to be 1 and re-derive the covariance matrix using the original jackknife error. This assures that the final covariance matrix for sample variance has diagonal elements unchanged. 

Smoothing the covariance matrix reduces the noise within it such that we can ignore the contribution to the posterior in a Bayesian analysis. An alternative would be to jointly fit to the data and covariance, leading to a posterior with a non-Gaussian form, which depends on the prior placed on the covariance matrix \citep{SH-2016,Percival-2022}. We use the Gaussian kernel smoothing method as default and simply set the width of the kernel to be 1, but we test alternative corrections for the bias from an approximate covariance matrix in the Bayesian analysis including those from \cite{Hartlap_2007} and \cite{Percival-2022} in Appendix~\ref{appsec:covariance}. In addition, we also test the  proposal of \cite{Mohammad_2022} for a revised jackknife procedure. These different methods are shown to make negligible changes to our results.

\section{Results}\label{sec:result}

\subsection{Luminosity Dependence}\label{sec:luminosity}

\begin{figure*}
\begin{center}
\includegraphics[width=19cm]{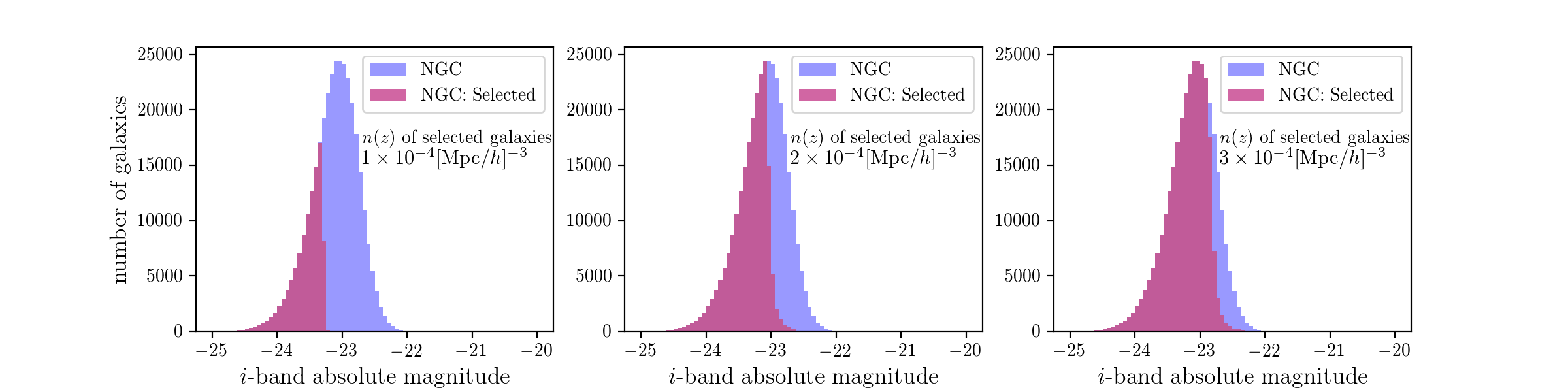}
\caption{Distribution of the i-band magnitude of BOSS galaxies. Galaxies selected by the number density threshold are marked in purple. For N1 (left) and N2 (middle) densities, the number density selection is close to a hard cutoff on luminosity, giving a more complete sample than N3 (right).}
\label{fig:maghist}
\end{center}
\end{figure*}

\begin{figure*}
\begin{center}
\includegraphics[width=17cm]{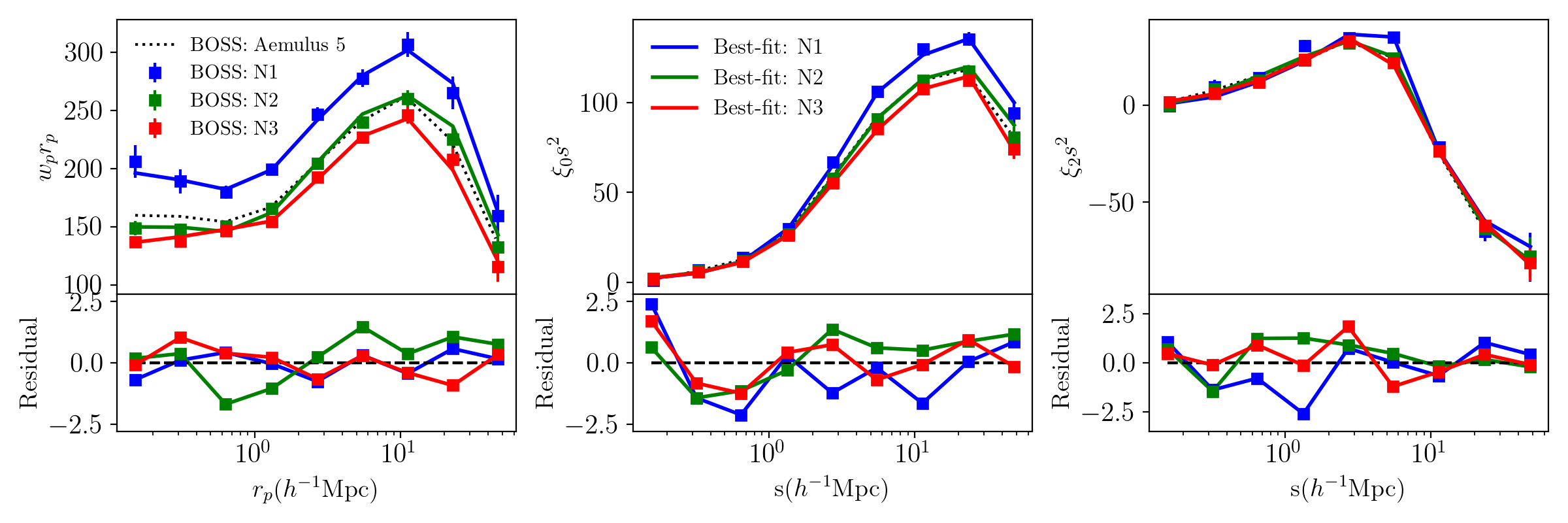}
\caption{Measurement of the correlation function, including $w_{p}$ (left), $\xi_{0}$ (middle) and $\xi_{2}$ (right) for BOSS galaxies selected by luminosity. The solid lines represent predictions from the best-fit model of our cosmology and galaxy bias model. For reference, the measurement from Z22 using BOSS NGC+SGC galaxies and N2 number density is shown by the dotted lines. The bottom panels show the residuals of our fit with respect to the measurement, normalized by the observed uncertainties. }
\label{fig:Bestfit_CF_luminosity}
\end{center}
\end{figure*}
 
As described in Section~\ref{sec:data}, we use the i-band magnitude to rank galaxies in thin redshift slices, selecting samples to three densities: N1, N2 and N3. As we increase the number density, more faint galaxies are selected to reduce the average brightness of the sample, and the galaxy sample with lower number density is a subset of the sample with higher number density, i.e. N1$\subset$N2$\subset$N3. Note that N2 selection is identical to Z22 with $0.48<z<0.62$. In Figure~\ref{fig:maghist}, we present the distribution of the i-band magnitude with different number densities and compare with the BOSS NGC sample. This shows that our number density selection is close to a hard cutoff of magnitude, especially with lower number density. The N3 subsample is close to the original BOSS galaxies with only a small fraction of faint galaxies removed. A more detailed examination of this incompleteness and its impact on the clustering analysis is necessary. However earlier works show that this incompleteness of BOSS galaxies doesn't have a significant impact on the clustering measurement (\citealt{Leauthaud_2017, Tinker_2017a}) and thus change the result of this paper. 

Figure~\ref{fig:Bestfit_CF_luminosity} displays the clustering measurements of the three subsamples, for $w_{p}, \xi_{0}$ and $\xi_{2}$ respectively. For reference, we also show the measurement from Z22 with N2 number density but for NGC+SGC. The consistency between NGC+SGC (dotted black) and NGC only (solid green) shows that galaxies in the NGC and SGC have consistent clustering amplitude with our selection algorithm. The dependence of clustering amplitude on the sample selection is in agreement with our expectations. The lowest number density N1 leads to the brightest galaxy sample and highest clustering amplitude. As fainter galaxies are included, the clustering amplitude decreases, since brighter galaxies preferentially live in more massive halos with higher large scale bias. We also plot the best-fit 2PCF predicted from the emulator compared with the BOSS measurements. The bottom panel shows the residual normalized by the observed jackknife uncertainty showing consistency from non-linear to linear scales for all three number densities. 

\begin{figure}
\begin{center}
\includegraphics[width=8cm]{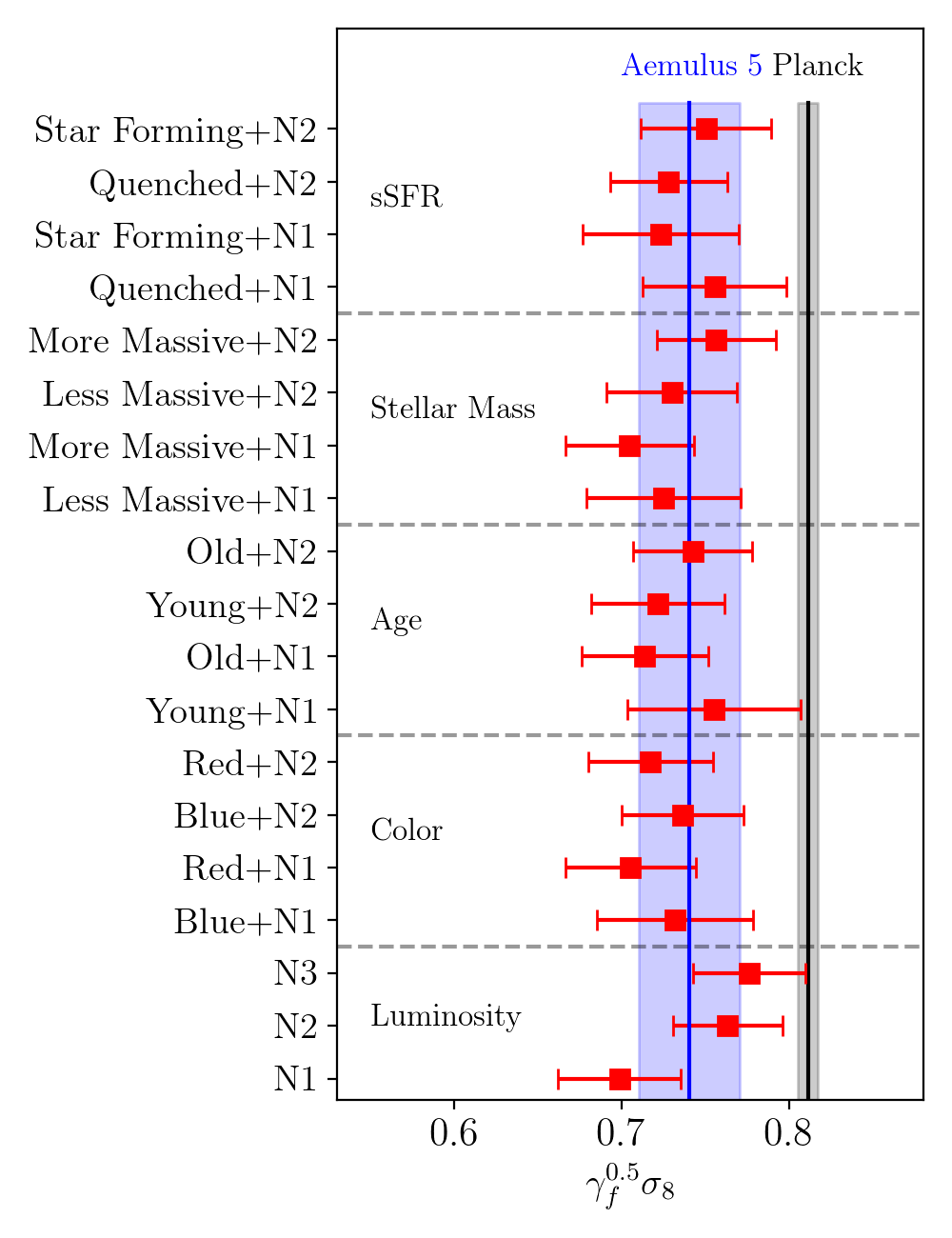}
\caption{Measurement of $\gamma_{f}^{0.5}\sigma_{8}$ from our analysis by splitting BOSS galaxies into different subsamples. For reference, we also show the result from Z22 (blue vertical shaded area) and Planck (grey vertical shaded area).}
\label{fig:f05sigma}
\end{center}
\end{figure}

We extract the measurement of the growth rate from the posterior distributions and express the result as $\gamma_{f}^{0.5}\sigma_{8}$ as shown in the bottom panel of Figure~\ref{fig:f05sigma}, where $\gamma_{f}$ is the velocity scaling parameter of dark matter halos (\citealt{Reid_2014, Zhai_2019}). The result from Z22 is also shown for comparison. Since our current analysis only uses the NGC, the uncertainty is slightly larger due to the smaller volume. We find that our galaxy samples with three different luminosities give roughly consistent measurements of the growth rate, with weak evidence that brighter galaxies give a slightly lower $f\sigma_{8}$, i.e. more offset compared with \cite{Planck_2020}. In addition, we note that the offset between results from the N1 and N2 subsamples is larger than might be expected given the fact that these two subsamples are correlated. In order to understand the discrepancy, we perform a consistency test. Starting from a cosmological model close to the Z22 best-fit, we generate multiple sets of correlated noise for N1 and N2 clustering measurements around this model, using the covariance matrix to produce realistic realizations of noise. We fit these mock data using the same pipeline as analyzing the data and get the cosmological constraints. From the results, we find that the discrepancy between N1 and N2 constraint could be related to the modeling of the emulator inaccuracies.
%likely caused by the emulator inaccuracies rather than the sample variance. 
In particular, we generate 10 realizations for this test. Without the emulator error, none of them has an offset larger than $1\sigma$ between N1 and N2 results, but 1 mock yields a $2\sigma$ discrepancy when the emulator error is added. At the scales in our analysis, the contribution from the emulator error to the final covariance matrix is comparable to the sample variance. We note that our estimate of the emulator inaccuracies is from the Aemulus test simulations and does not take into account correlation between different samples, a consequence of the low number of test samples available. A correlation is expected and would increase the discrepancy between the N1 and N2 subsamples.

Another feature in the constraints is that the uncertainty in the results from the N1 subsample is only slightly larger than that for the N2 subsample, e.g. in Figure \ref{fig:f05sigma}, the uncertainty in the growth-related parameters from the N1 subsample is only about 20\% larger than that from N2. Given their difference in number density, this scaling is less than expected for an analysis that is shot-noise dominated (\citealt{Dawson_2022}). There are multiple reasons for this. The first is that the emulator error in the likelihood analysis is not negligible compared with sample variance and shot noise, and dilutes the difference of the total covariance. We assume that the emulator error is uncorrelated between samples. Second, most of the cosmological constraining power in the current clustering analysis is from scales $\sim$ a few $h^{-1}$Mpc, which marks a transition from linear to non-linear scales. In terms of variance, this scale is only slightly lower than the expected transition from sample variance domination to shot-noise domination, which is quite a gradual change. Therefore the change of number density may not perfectly scale with the change of constraining power as expected for a shot-noise dominated sample. This is linked to the information available at different scales in the standard two-point statistics (\citealt{Storey-Fisher_2022}); see for instance the scale-dependent analysis of Figure 10 in Z22 and Figure 8 in \cite{Zhai_2019}.

Due to the increased uncertainty, the tension between the brightest galaxy sample (N1) and Planck is at a similar level to that seen in Z22. In order to evaluate the internal (in)consistency of the BOSS galaxies, we define a metric for the tension of $f\sigma_{8}$ from different galaxy samples
\begin{equation}\label{eq:tension}
    T = \frac{|F_{A5} - F_{N}|}{\sigma},
\end{equation}
where $F_{A5}$ is the result from Z22, $F_{N}$ is the result for a selected galaxy sample in the current analysis. A more conventional metric would be to use $\sqrt{\sigma_{1}^{2}+\sigma_{2}^{2}}$ in the denominator for two experiments. However, our galaxy samples are not fully independent, therefore we quote result as T=1.2 (0.9)$\sigma$ for N1 with $\sigma=\sigma_{f\sigma_{8, A5}}$ ($\sigma_{f\sigma_{8, N}}$). Similarly, T=0.8 (0.7)$\sigma$ for N2, and T=1.1 (1.2)$\sigma$ for N3. Note that this is a more severe evaluation for the consistency of different galaxy samples. The overall result shows that selecting galaxies with different luminosities won't change our measurement of $f\sigma_{8}$ by more than 1.2$\sigma$.

The fits also allow us to interpret the clustering measurement in the HOD framework after marginalizing over cosmology. The top panel of Figure~\ref{fig:HOD_all} shows the halo occupation distribution for galaxies with different luminosities. In order to isolate the effect due to different luminosities, we compute the probability that a galaxy in our sample is hosted by a dark matter halo of mass M$_{\text{halo}}$ shown in the bottom panels. One interesting feature from this figure is that all the satellites live in roughly the same range of halo mass $\sim10^{13.7}[h^{-1}M_{\odot}]$, while the overall population is significantly dominated by centrals with $f_{\text{sat}}\sim10\%$ or less, which matches the expectation that brighter galaxies live in more massive halos and that the sample is dominated by Luminous Red Galaxies. 

In Table~\ref{tab:constraint}, we report the measurement of the other galaxy bias parameters. Brighter galaxies are dominated by centrals, with a satellite fraction of only $\sim5\%$. For fainter galaxies, the satellite fraction can increase up to 9\%. This trend is consistent with that previously recovered from the SDSS Main Galaxy Sample (\citealt{Zehavi_2011}) and luminous red galaxies (\citealt{Zheng_2009, Guo_2014}) fitting different models. Since our number density selection is equivalent to a threshold on luminosity, i.e. a lower number density corresponds to a brighter galaxy sample, the result implies a correlation between number density and satellite fraction. This is in agreement with the results in literature, see e.g. Figure 5 of \cite{Guo_2014}. In addition, we find that brighter galaxies have higher velocity bias for both centrals and satellites, with respect to their host halos. This can enhance the RSD and reduce the inferred growth rate of structure predicted by a given cosmology. Another luminosity or number density dependent parameter is the assembly bias parameter. Brighter galaxies can have a mild positive value of $f_{\text{env}}$ while fainter galaxies reduce this value until it is negative. This implies that bright galaxies in over-dense regions live in more massive halos than galaxies in under-dense regions, while faint galaxies have the opposite preference. A thorough investigation on the environment and its impact on the galaxy formation efficiency will better reveal the detailed physics but is beyond the scope of this paper. For the remaining concentration parameters affecting the satellite distribution $\eta_{\text{con}}$ and $\gamma_{f}$, we do not find significant dependence on luminosity. All galaxy samples show similar constraints on $\eta_{\text{con}}<1$, indicating the spatial distribution of satellites is less concentrated than the NFW profile but the deviation is not significant. The velocity scaling parameter $\gamma_{f}$ is slightly less than 1 for different luminosities but is not in significant tension with $w$CDM+GR predictions.

\begin{figure*}
\begin{center}
\includegraphics[width=17.5cm]{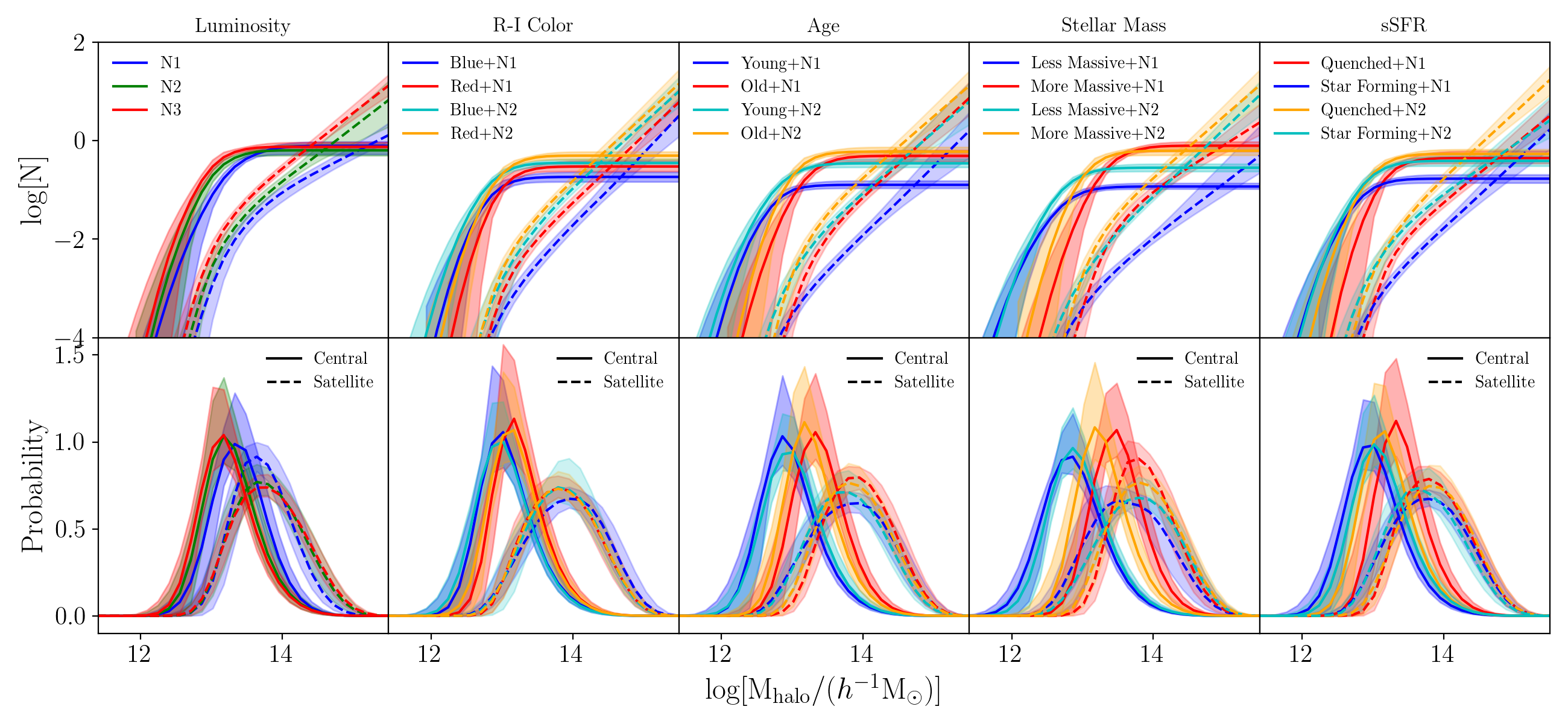}
\caption{Top: halo occupation distribution of halos as a function of halo mass for galaxy subsamples. Solid and dashed lines represent the mean occupancy for central and satellite galaxies respectively. Bottom: Probability distribution of galaxies as a function of host halo mass. The columns correspond to selections by luminosity (first column), color (second column), age (third column), stellar mass (fourth column) and sSFR (fifth column). }
\label{fig:HOD_all}
\end{center}
\end{figure*}

\subsection{Color Dependence} \label{sec:color}
\begin{figure*}
\begin{center}
\includegraphics[width=18cm]{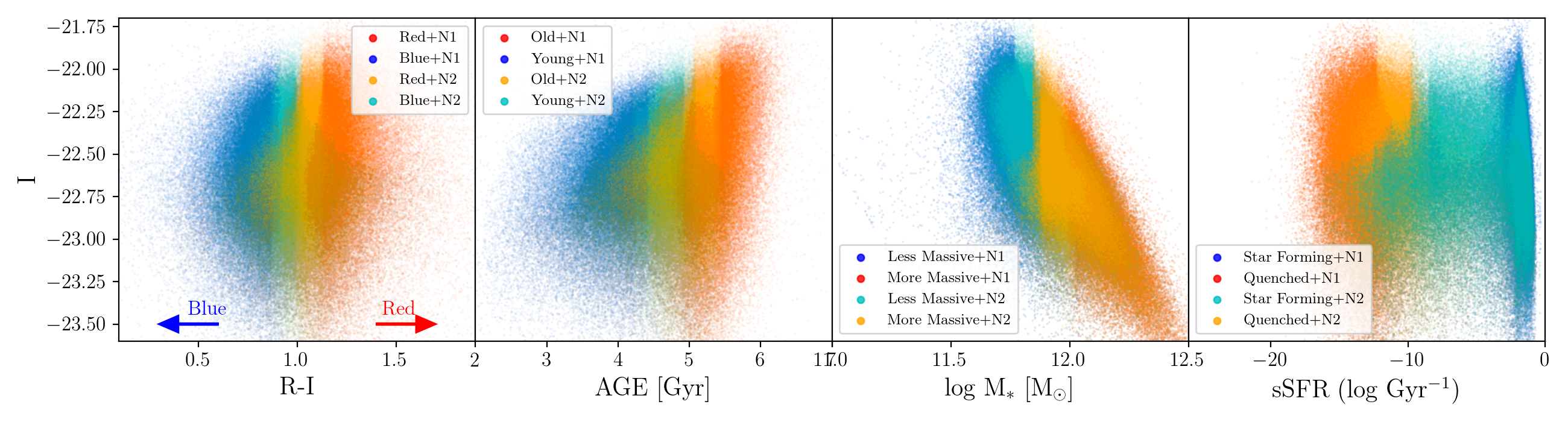}
\caption{Scatter plot for I-band magnitude and derived properties (color, age, stellar mass and sSFR) of the BOSS galaxies with $0.48<z<0.62$ from one of the Granada VACs (\citealt{Salpeter_1955} IMF + wide prior on galaxy age for the star formation model + dust attenuation). We select galaxies from the extremes such that the final galaxy sample can reach number densities of N1 and N2 respectively. Note that our color definition is applied for the number density selection algorithm, other definitions involving multiple bands are also used in literature, e.g. \protect\cite{Zehavi_2011, Guo_2013} and \protect\cite{Ross_2014}.}
\label{fig:scatter_mag_all}
\end{center}
\end{figure*}

\begin{figure*}
\begin{center}
\includegraphics[width=17cm]{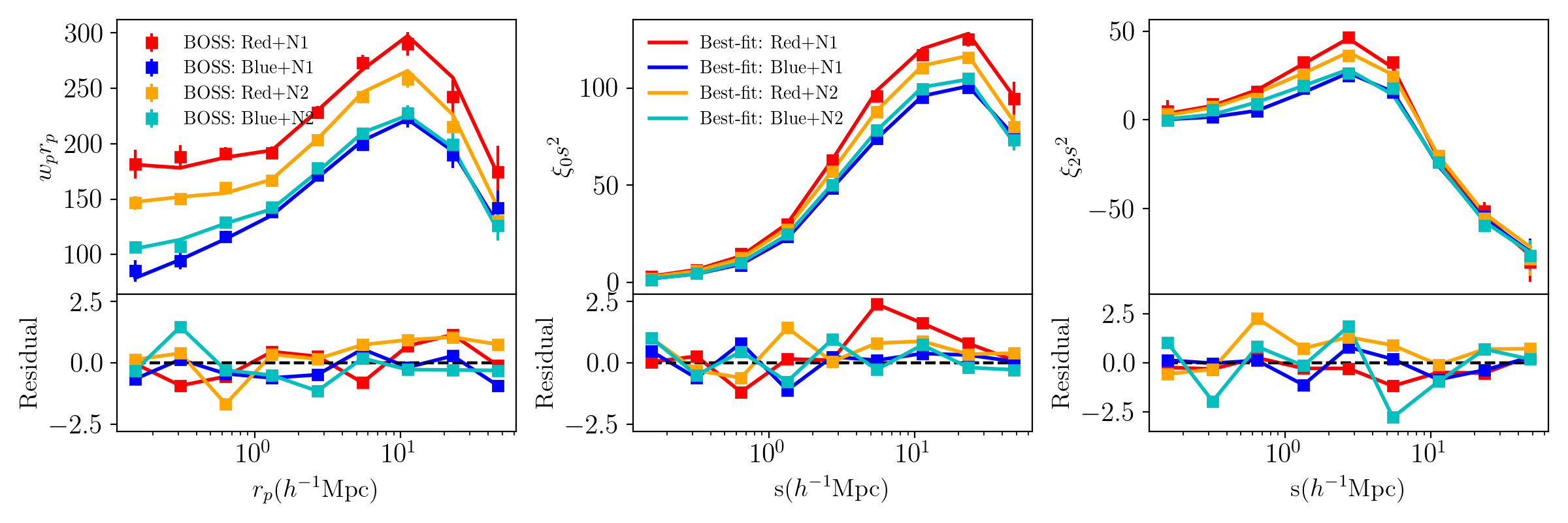}
\caption{Same as Figure~\ref{fig:Bestfit_CF_luminosity} but for color-dependent analysis.}
\label{fig:Bestfit_CF_color}
\end{center}
\end{figure*}

The definition of galaxy color requires magnitudes in multiple bands, and the split into red and blue galaxies involves some combination of these, see e.g. \cite{Zehavi_2011}. In this analysis, we choose a split in R-I as in \cite{Ross_2014} and \cite{Guo_2014} as the color proxy to create galaxy samples. In the first panel of Figure~\ref{fig:scatter_mag_all}, we show a scatter plot of the I-magnitude and R-I color for the BOSS galaxies, color coded by the number density selection. Applying the N1 number density produces the ``reddest" and ``bluest" galaxy samples, while the N2 samples are more similar with each other. At redshift above $\sim$0.57, all galaxies can be selected. And at lower redshift, the fraction of selected galaxies is still high and the clustering may not be distinct enough. The figure seems to imply that the threshold of a constant number density is close to a cut for galaxy color, especially for fainter galaxies. 

We create color-dependent samples using the eight Granada VACs, for variations of the priors on star formation scenarios, dust and initial mass functions. Although the Granada templates provide different sets of properties for individual galaxies, potentially leading to a complicated set of model fits, we note that the clustering measurement is insensitive to the choice of the template parameters, since our selection algorithm based on rank-ordering the galaxy color reduces this impact. The measurements of galaxy correlation function from these catalogs are presented in Appendix~\ref{appsec:CF}. 

Given the consistency of the VACs, we choose to present results from the template combination including a wide prior on galaxy lifetime for star formation + dust + \cite{Salpeter_1955} initial mass function, throughout our analyses. We also use the clustering measurement from this VAC for the tests of other galaxy properties in subsequent sections. 

In Figure~\ref{fig:Bestfit_CF_color}, we present the best-fit results and residuals, which show that our model is able to provide good fits to the color-selected galaxies. With these results, we first examine the measurements of growth rate and show the constraint in Figure~\ref{fig:f05sigma}. It is clear that both red and blue subsamples give consistent constraints on the growth rate of structure. Using the metric of Equation~(\ref{eq:tension}), we find that all the four subsamples are in agreement with Z22 and the highest tension is less than $1.3\sigma$. This provides further evidence that the cosmological analysis is robust under different galaxy properties. 

Within the HOD framework, the second column of Figure~\ref{fig:HOD_all} shows the halo occupation of the color-selected samples and the probability distribution of the halo mass hosting the galaxies. Red galaxies reside in more massive halos than blue galaxies and thus have higher bias. Constraints on other galaxy bias parameters are summarized in Table~\ref{tab:constraint}. We can see that the red galaxies have more satellites than blue galaxies on average, consistent with $w_{p}$ being enhanced at small scales, where the higher satellite fraction gives more 1-halo pairs. The combined result is consistent with the standard picture (e.g. \citealt{Guo_2014}) that the satellite fraction mainly affects small scale clustering, while the average mass scales of dark matter halos can determine the overall clustering amplitude.  In addition, we find that $f_{\text{max}}$ for more extreme samples (N1) is much lower than for luminosity selected samples. Adding less-extreme galaxies can significantly increase $f_{\text{max}}$. We notice that this feature also exists for galaxy samples selected by other properties, implying that central galaxies with low number density can have occupancy different from the asymptotic value modeled by the error function in the HOD formalism. In addition, \cite{Chapman_2021} showed that a low value of $f_{\text{max}}$ can lead to a lower value of $f\sigma_{8}$. However, our result seems to show that the color-selected sample is less affected by this parameter for cosmological inference. Among these galaxy bias parameters, only $\eta_{\text{vs}}$ has some dependency on galaxy color, i.e. blue satellites have lower velocity bias than red satellites. However, its impact on cosmological inference is reduced due to the low satellite fraction. 

\subsection{Age Dependence}

\begin{figure*}
\begin{center}
\includegraphics[width=17cm]{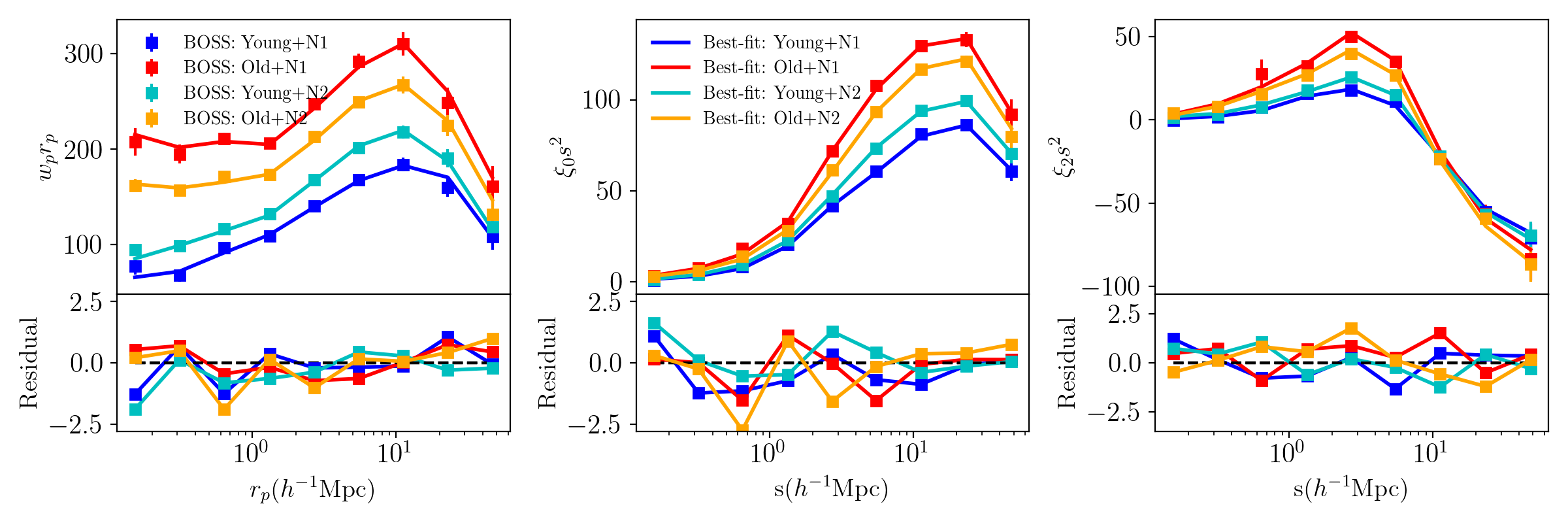}
\caption{Same as Figure~\ref{fig:Bestfit_CF_luminosity} but for age dependent analysis.}
\label{fig:Bestfit_CF_age}
\end{center}
\end{figure*}

The second panel of Figure~\ref{fig:scatter_mag_all} shows a scatter plot of I-magnitude and age for the BOSS galaxies. The age parameter corresponds to the mass-weighted average age of the stellar population from the Granada VACs. Our sample selection algorithm defines four subsamples of young and old galaxies, in a similar fashion to the color-selected sample. Figure~\ref{fig:Bestfit_CF_age} shows the clustering measurements and our best-fit model prediction. The result shows that older galaxies are more strongly clustered than younger galaxies, consistent with results for galaxies at lower redshift (\citealt{Zehavi_2011}). Combined with the measurement of color-selected galaxies, this matches the standard correlation between galaxy color and age, i.e. red galaxies are generally older than their blue counterparts. Our emulator-based model with varying cosmology and galaxy bias is able to provide a reasonable fit to the observed clustering measurements for all bins. In Figure~\ref{fig:f05sigma}, we present the measurement of growth rate parameters from these age-selected samples. The results indicate that the measurements are consistent with each other. Compared with Z22, the tension is less than T=1.1 (0.8)$\sigma$. 

In the third column of Figure~\ref{fig:HOD_all} and in Table~\ref{tab:constraint}, we present the constraints on the HOD and galaxy bias parameters for the age-selected galaxy samples. Compared with the color selected sample, the age-selected galaxies have more distinct populations in halo mass. The central galaxies of the youngest and oldest samples live in halo masses with a difference of $\sim0.4$ dex, with the older galaxies residing in more massive halos. In comparison, the reddest and bluest galaxies live in halos with a mass difference of about 0.2 dex. Due to the correlation between galaxy age and color, the constraint on the HOD parameters shows a similar behavior as in the second column of Figure~\ref{fig:HOD_all} showing the effect of color. Both young and blue galaxies have low satellite fractions and $f_{\text{max}}$, implying that they are rarely observed in the most massive halos. The other galaxy bias parameters are similar to each other showing no significant dependence on galaxy age. The most extreme is the measurement of $f_{\text{env}}$ from the youngest galaxies, but its offset compared with other samples is not significant. Given the correlation between luminosity, color and age. It's likely that the positive value of $f_{\text{env}}$ is affected by the degeneracy with the satellite fraction and $f_{\text{max}}$.

\subsection{Stellar Mass Dependence}

\begin{figure*}
\begin{center}
\includegraphics[width=17cm]{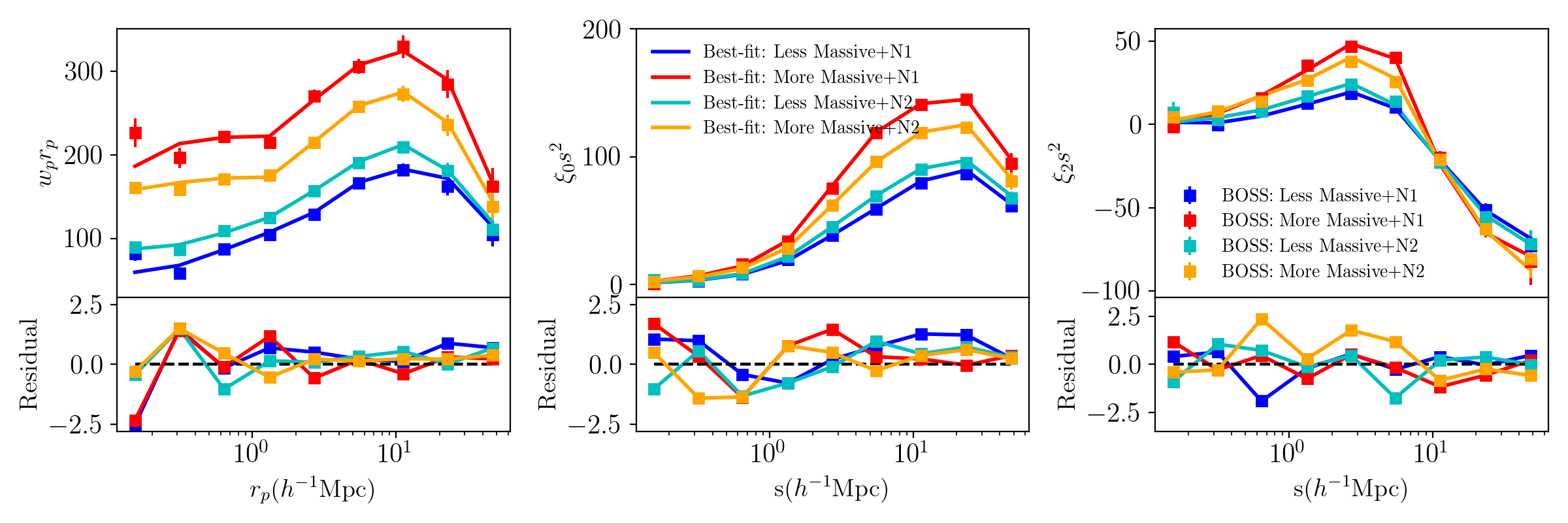}
\caption{Same as Figure~\ref{fig:Bestfit_CF_luminosity} but for stellar mass dependent analysis.}
\label{fig:Bestfit_CF_mstar}
\end{center}
\end{figure*}

The third panel of Figure~\ref{fig:scatter_mag_all} shows a scatter plot of I-magnitude and stellar mass, which clearly shows the strong correlation between galaxy brightness and stellar mass. We can consider that our sample selection starts from the two horizontal ends of this figure moving inwards until the required number density is reached. The final cuts in galaxy selection are sharp in stellar mass as would be expected for a volume limited sample, similar to what we found for the color selection in Section~\ref{sec:color}. We present the clustering measurements and the best-fit results in Figure~\ref{fig:Bestfit_CF_mstar}. As expected, the most massive galaxies have the highest clustering amplitude (\citealt{Li_2006a}). The difference in clustering amplitude between the most massive and least massive is higher than selections based on other galaxy properties. 

We show the measurement of growth rate compared with Z22 in Figure~\ref{fig:f05sigma}. The highest tension is observed for the most massive galaxy sample with T=1.7 (1.2) $\sigma$, indicating that the cosmological constraint under this stellar mass split is still robust. The HOD and galaxy bias parameters for these galaxy samples are shown in the fourth column of Figure~\ref{fig:HOD_all} and in Table~\ref{tab:constraint}. The correlation between stellar mass and halo mass is clearly seen in the bottom panel of Figure~\ref{fig:HOD_all}. The most and least massive galaxies live in dark matter halos with a mass difference of up to 0.6 dex, covering a wider range of halo mass than the results inferred from color or age dependent clustering analysis. In addition, the change of $f_{\text{max}}$ due to stellar mass is similar to that we saw with galaxy age, matching the positive correlation between galaxy age and stellar mass, i.e. older galaxies have more time to accumulate gas for star formation activity and thus increase stellar mass. 

The galaxies with stellar mass below $10^{11.8}M_{\odot}$ have a higher velocity bias for centrals than more massive galaxies. This is not found in young or blue galaxies although the difference is not higher than $2\sigma$, implying some scatter between stellar mass and other galaxy properties at the less massive end. The other galaxy bias parameters do not have such a strong dependency on stellar mass. Note that the tiny uncertainty of $\eta_{\text{con}}$ for the most massive sample, which is caused by the closeness to the lower boundary of the prior range.

\subsection{sSFR Dependence}\label{sec:SP}

\begin{figure*}
\begin{center}
\includegraphics[width=17cm]{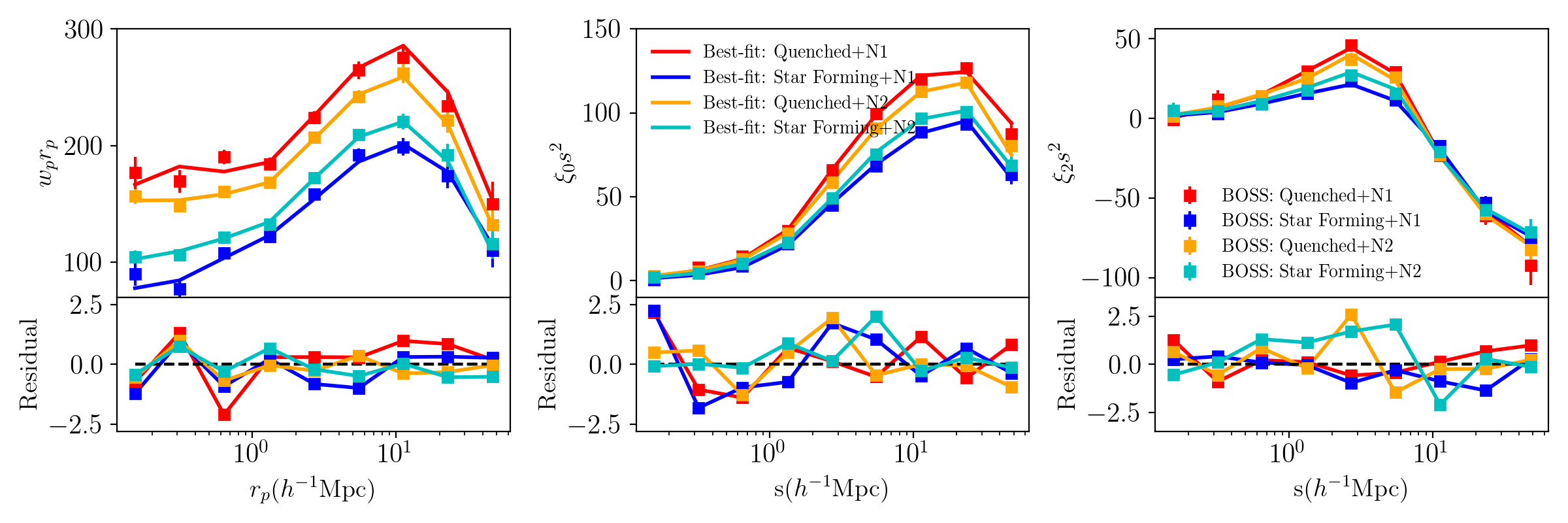}
\caption{Same as Figure~\ref{fig:Bestfit_CF_luminosity} but for sSFR dependent analysis.}
\label{fig:Bestfit_CF_ssfr}
\end{center}
\end{figure*}

Although correlated with color and age, the specific star formation rate (sSFR) is also provided in the catalogs, with which we can split the sample. In the fourth panel of  Figure~\ref{fig:scatter_mag_all}, we present a scatter plot of the I-magnitude against sSFR from the Granada VAC. We split the sample into star forming and quenched populations using the selection algorithm. Note that the term "star forming" or "quenched" only refers to their location in our selection, without actually meaning they are star forming or quenched. The figure shows a bimodal distribution for sSFR, i.e. galaxies are more concentrated at the star forming and quenched ends. However, this bimodal distribution is not significant on the color separation (Figure~\ref{fig:scatter_mag_all}), implying the scatter of the correlation between galaxy color and sSFR for BOSS galaxies. We note that this is also observed in earlier analysis of BOSS galaxies (\citealt{Ross_2014}), which combines r-i and i-magnitude to define red and blue galaxies. 

In Figure~\ref{fig:Bestfit_CF_ssfr}, we present the clustering measurements for the sSFR selected galaxy samples. Similar to previous sections, splits in sSFR give samples with very different clustering amplitudes. The quenched galaxies have higher clustering amplitude, consistent with red, old and more massive galaxies. From the fitting results, we present the best-fit growth rate parameters in Figure~\ref{fig:f05sigma}, which shows that the measurements are in agreement with Z22. We use Equation~(\ref{eq:tension}) to measure the consistency between them and find that the highest tension is from the quenched+N1 subsample with T=0.7 (0.4) $\sigma$. We present the HOD and constraints on the galaxy bias parameters in the fifth column of Figure~\ref{fig:HOD_all} and in Table~\ref{tab:constraint}. The HOD is similar to our previous analyses in that the central galaxies with different sSFR live in different halo mass. Star forming galaxies live in lower mass halos on average. In addition, the satellite fraction $f_{\text{sat}}$ and $f_{\text{max}}$ of these star forming galaxies is lower than quenched galaxies. Another interesting feature is that there is a correlation between sSFR and velocity bias for centrals. The star forming centrals have higher velocity bias than the quenched galaxies, although we should keep in mind that the uncertainty is large and therefore the dependence is not significant. It is possible that these galaxies have experienced merger or entered the gravitational well recently. The other galaxy bias parameters do not show significant dependency on sSFR, for instance the velocity bias for satellites and $f_{\text{env}}$.

\subsection{Dependence on SP Parameters} \label{sec:result2}

\begin{figure*}
\begin{center}
\includegraphics[width=17cm]{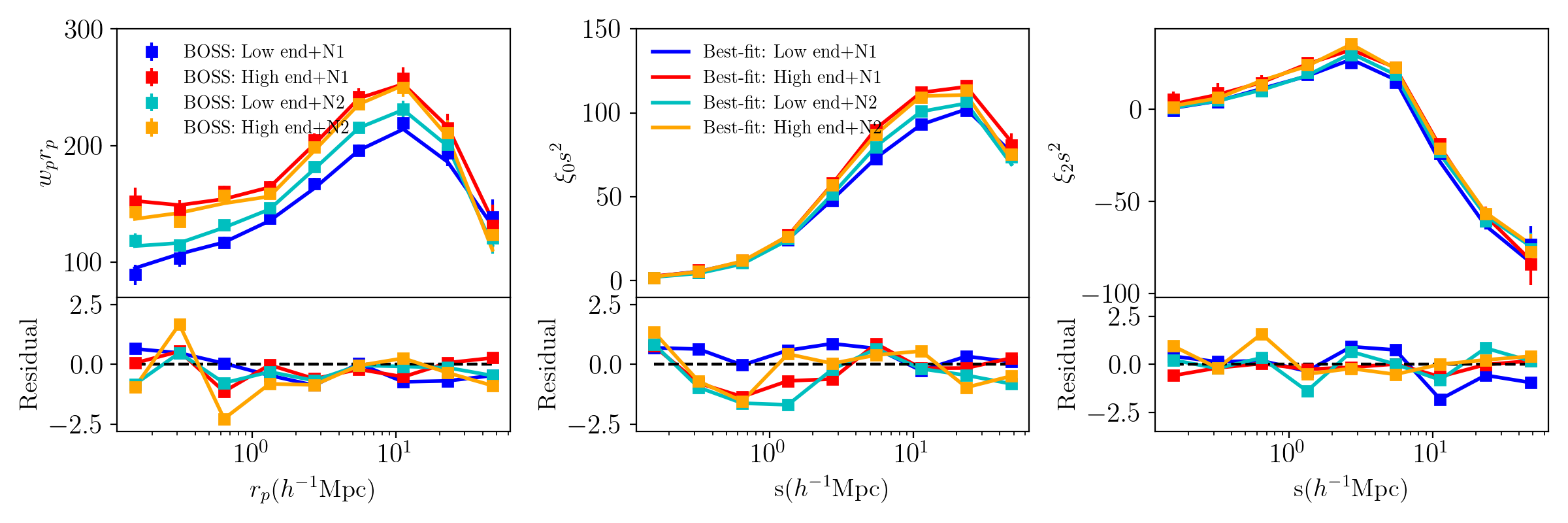}
\includegraphics[width=17cm]{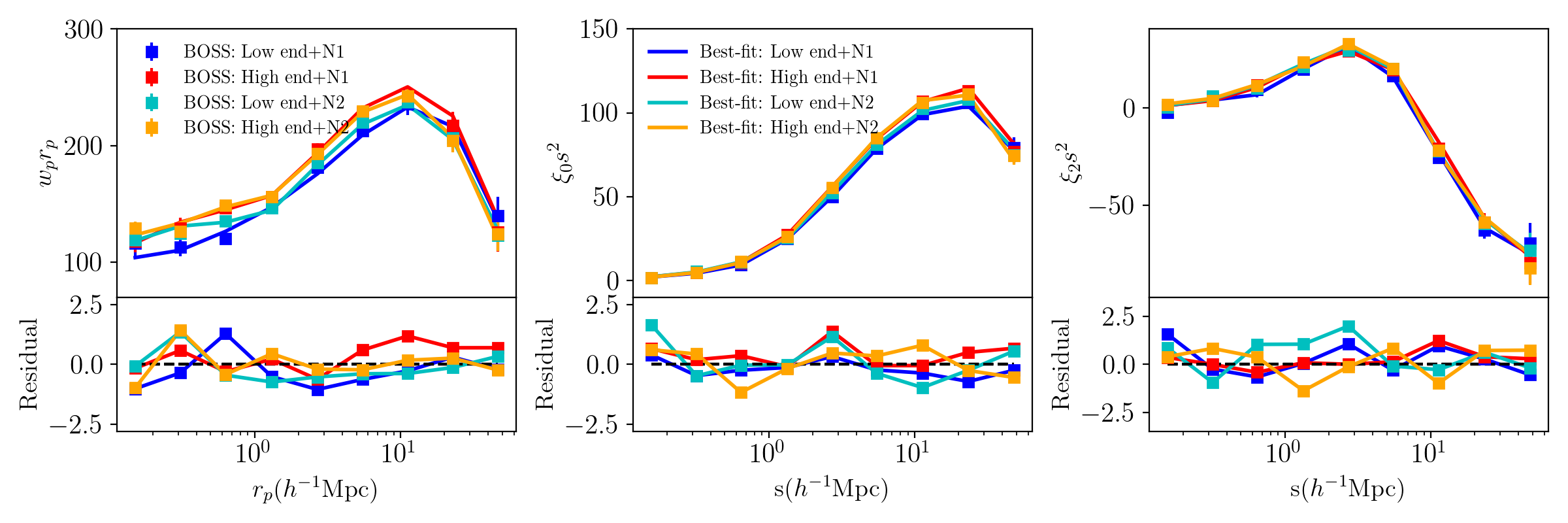}
\caption{Same as Figure~\ref{fig:Bestfit_CF_luminosity} but for SP parameter dependent analysis. Top is for look-back formation time and bottom is for metallicity.}
\label{fig:Bestfit_CF_lbf_metal}
\end{center}
\end{figure*}

\begin{figure*}
\begin{center}
\includegraphics[width=8cm]{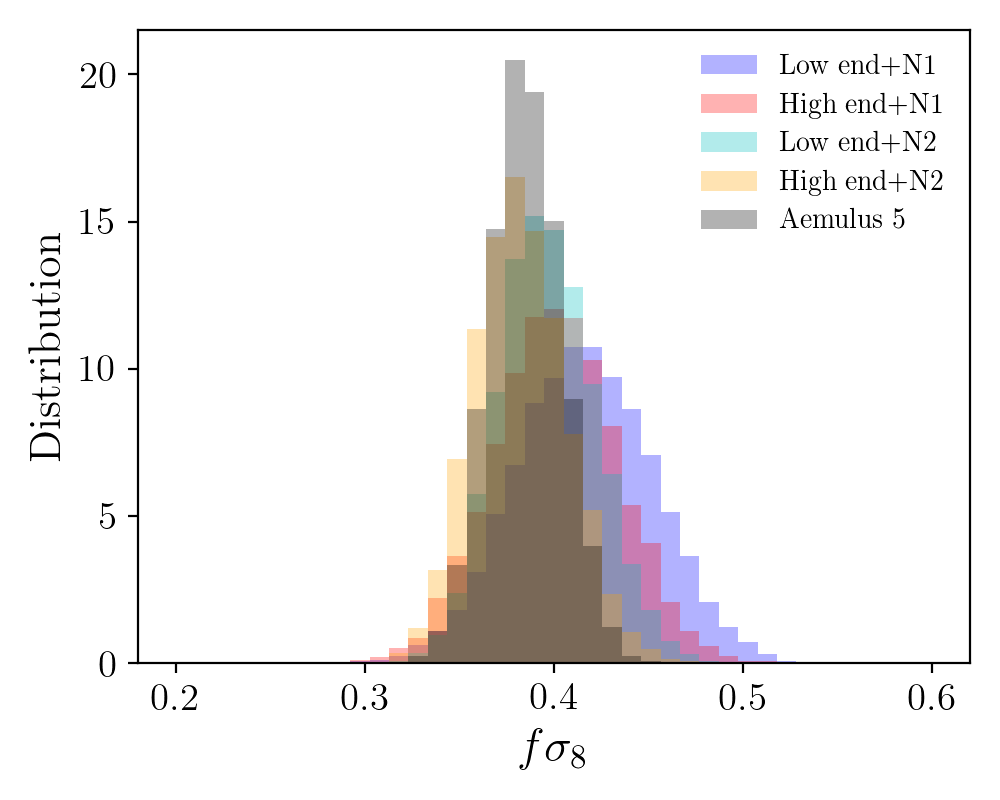}
\includegraphics[width=8cm]{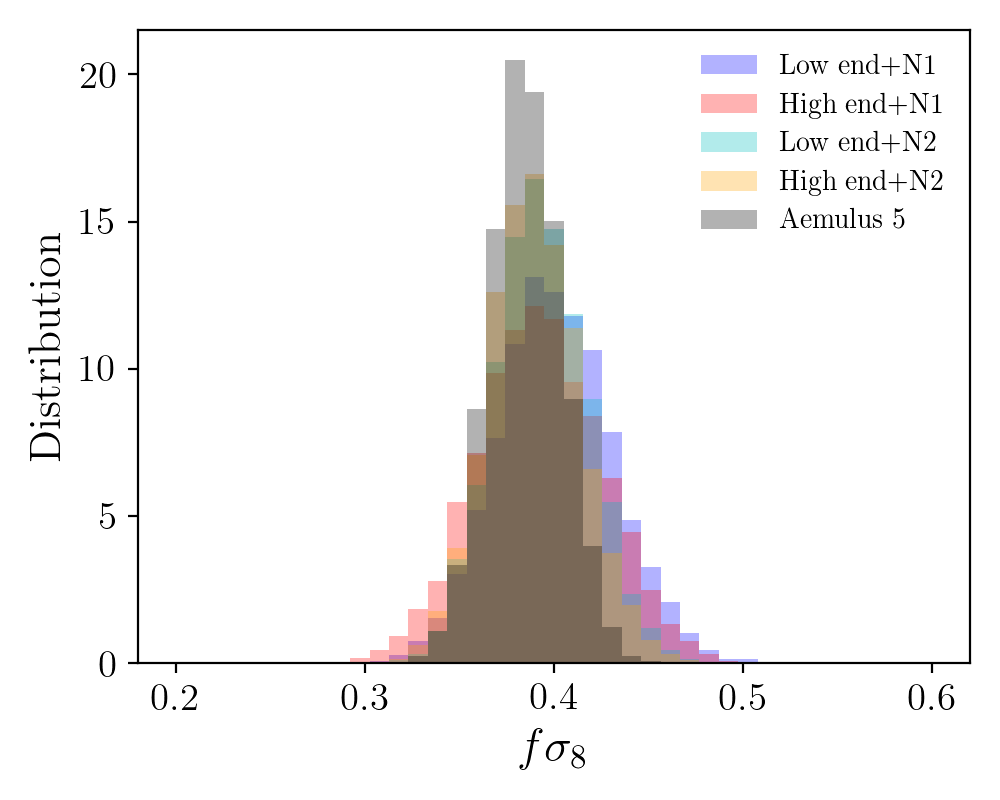}
\caption{Measurement of $f\sigma_{8}$ for SP parameter dependent analysis. Left is for look-back formation time and right is for metallicity.}
\label{fig:fsigma_lbf_metal}
\end{center}
\end{figure*}

In the previous sections we have analysed the clustering measurements of galaxy samples defined by different core galaxy properties. The Granada VAC also reports a number of secondary parameters from the fitting of SP model including look-back formation time, metallicity, dust attenuation around young and old stars, and star formation history e-folding time. We run additional tests using these parameters in this section. In particular, we choose look-back formation time and metallicity to construct galaxy subsamples using the selection algorithm described in Section~\ref{sec:data}. These parameters can be correlated with galaxy parameters tested in the previous sections, for instance the correlation between galaxy color and metallicity etc. Therefore the test with these SP parameters serves as a sanity check of our modeling of cosmology and galaxy bias. 

After constructing galaxy samples, we measure the correlation function of the subsamples as shown in Figure~\ref{fig:Bestfit_CF_lbf_metal}. The result indicates the dependence of clustering on look-back formation time. Galaxies with longer formation times cluster more strongly, which is in agreement with galaxy age but the dependence is less significant. The clustering dependence on metallicity is weak, although the tendency is consistent with the expectation that redder galaxies have higher metallicity. However, the difference between the clustering amplitudes of the two subsamples with lowest and highest metallicities is not as high as for the color or age selected samples. 

Given these measurements, we project the constraint on the model parameters and show the best-fit correlation functions in Figure~\ref{fig:Bestfit_CF_lbf_metal}. This shows that our model can still provide a reasonable fit to the clustering measurements, even if the galaxies are split into subsamples using less commonly analysed parameters. Figure~\ref{fig:fsigma_lbf_metal} displays the measurement of $f\sigma_{8}$ using galaxy samples selected by these SP parameters. We can see that the distribution is also consistent with the previous results. For look-back formation time, we find that the agreement is T=0.9 (0.5) $\sigma$ or better. For metallicity, it is T=0.3 (0.3) $\sigma$ or better. For both parameters we see consistent measurements of growth rate as Z22. Then we measure the HOD and obtain constraints on galaxy bias parameters. The impact of SP parameters on galaxy clustering is correlated with host halo mass. The galaxy samples with longer formation time preferentially live in more massive halos, indicating these galaxies are also redder, older, less active in star forming and more massive based on our previous analysis. For metallicity selected samples, since their clustering amplitude is closer to each other, the HOD is  similar for the centrals and the amplitude is dominated by the galaxy number density. The central galaxies live in halos with mass at $\sim10^{13}M_{\odot}$ for the entire range of metallicity parameter considered. The galaxy bias parameters for these formation time selected or metallicity selected samples are not significantly different. From these results, we can see that splitting galaxies by look-back formation time is similar to previous physical parameters. However, splitting by SP metallicity is close to a random downsampling, in terms of clustering amplitude, shape and inferred galaxy bias parameters.

\section{Discussion}\label{sec:discussion}

We have performed an analysis of the clustering of subsamples drawn from the BOSS NGC galaxy sample, defined by physical properties including luminosity, color, age, stellar mass and sSFR. We find consistent constraints on the cosmological parameters for all of the catalogues tested. Due to the reduced volume of using only the NGC and the reduced density from sub-sampling, the measurement of $f\sigma_{8}$ from these samples are less accurate as Z22, but we find that the consistency is better than $2\sigma$ for all sub-samples. The overall measurement of $f\sigma_{8}$ is in agreement with the other works using different simulations and methodologies, including \cite{Lange_2021, Chapman_2021, Yuan_2022}. We refer the readers to Z22 for a thorough comparison.

Assuming a Gaussian likelihood, the goodness-of-fit in our analysis can be evaluated by the $\chi^{2}$ value, given the number of data points and degrees of freedom of the model. The last column of Table \ref{tab:constraint} presents $\chi^{2}$ value for our tests. We note that for subsamples with number density of N2 and N3, all the tests present reasonable fits, i.e. the reduced $\chi^{2}$ is around unity. However, some galaxy subsamples with the lowest density N1 show worse fits, including the brightest galaxies and subsamples split by stellar mass and sSFR. Based on the best-fit correlation function, it seems that this high $\chi^{2}$ is dominated by one to two data points in each data vector. For instance, in Figure \ref{fig:Bestfit_CF_luminosity}, the solid blue lines show the most significant offset of monopole and quadrupole around $1~h^{-1}$Mpc. In Figure \ref{fig:Bestfit_CF_mstar}, the N1 galaxies (most and least massive) have the highest offset for $w_{p}$ at the smallest scale, while in Figure \ref{fig:Bestfit_CF_ssfr}, the deviation is dominated by the innermost data point from the monopole measurement for galaxies with highest and lowest sSFR. We do not find that the bad $\chi^{2}$ is caused by a systematic offset over a wide range of scale, and removing these problematic data points can significantly improve the goodness of fit. Therefore it's likely that there is noise in the measurement that is not included in the covariance matrix, which contributes to the $\chi^{2}$ given the lowest density of the galaxy samples.

To test the possibility that noise in the correlation matrix from the Jackknife algorithm is affecting our results, we have artificially replaced "noisy" correlation matrices by "less noisy" ones for some fits. In particular, we choose the galaxy subsample with large $\chi^{2}$, and change the correlation matrix in the likelihood with another correlation matrix from a galaxy subsample with a smaller $\chi^{2}$. Note that we keep the diagonal elements in the covariance matrix unchanged. We find that $\chi^{2}$ is only changed by a negligible amount. This shows that the noise in the correlation matrix doesn't bias the analysis, through either their intrinsic scatter or the Gaussian smoothing of this. 

We also note that not all the parameters in our model contribute equally in the measurement of $f\sigma_{8}$. Some of the parameters can be more important in determining the clustering shape and amplitude than the others. Therefore the number of parameters in the model may not be an ideal metric for the degrees of freedom. See a similar discussion in \cite{Lange_2021}. However, from a more general point of view, the parameters that are not critical for the measurement of $f\sigma_{8}$ can still change the model, although in a more implicit way.

In addition to the cosmological constraints, we have investigated the distribution of galaxy bias parameters for different subsamples. We find the standard result that there is a similar dependence in samples split by galaxy luminosity, color, age, stellar mass and sSFR results caused by intrinsic correlations between these properties. In the HOD formalism, this can be described through their correlation with dark matter halo mass. Although we extend the basic mass-only dependent HOD model by considering galaxy kinematics and assembly bias, we find that the key information is well captured by halo mass. However, we should note that the scatter exists between these parameters and halo mass due to the complicated processes governing galaxy formation including those dependent on baryonic physics (\citealt{Tinker_2017a, Tinker_2017b}). For instance, the less massive galaxies have a central velocity bias that is slightly different from young or blue galaxies, albeit with significant uncertainty on these results.

We also note that the satellite fraction is dependent on luminosity and color, i.e. fainter or redder galaxies prefer higher satellite fractions. For centrals, the most luminous galaxies have a large-scale bias (dependent on halo mass) that increases with luminosity. For satellites, there are less clear relationships between halo mass and luminosity, such that the clustering amplitude of redder galaxies will depend on the satellite fraction. An earlier study by \cite{Cresswell_2009} showed that faint red galaxies can be more biased than those at higher luminosities, suggesting that they are satellites living in more massive halos and thus probe more biased density field. The impact of satellite fraction on clustering amplitude may be even more complicated than this, since the satellites can live in less massive subhalos but their host halo can be more massive. The balance between average halo mass for centrals and satellite fraction is the important factor.

The constraints on the galaxy kinematics shows that the BOSS galaxies need an additional velocity bias, especially for centrals regardless of how we select the sample \citep{Guo_2015}. The velocity bias parameter for centrals $\eta_{\text{vc}}$ is around 0.3, which is defined as the velocities of central galaxies with respect to their host halos, expressed in units of the velocity dispersion of the dark matter halos. This additional velocity can enhance the RSD effect through its degeneracy with cosmological parameters. Its positive value may imply that the velocity field predicted by GR may not fully capture the RSD effect at small scales, even we introduce the parameter $\gamma_{f}$ to make the model more flexible. One example is the luminosity dependent analysis in Section~\ref{sec:luminosity}, brighter galaxies that require a higher value of $\eta_{\text{vc}}$ (and $\eta_{\text{vs}}$) have a lower measurement of $f\sigma_{8}$. Clearly, the information about structure growth from the velocity field comes from both small scales where the clustering is dominated by the HOD, and large-scales where the clustering is dominated by the cosmological model. This indicates the necessity of accurate modeling of baryonic physics at small scales, as well as a better understanding of non-linear dynamics of dark matter velocity field \citep{Guo_2015b,Chapman_2023}.

\begin{figure}
\begin{center}
\includegraphics[width=8cm]{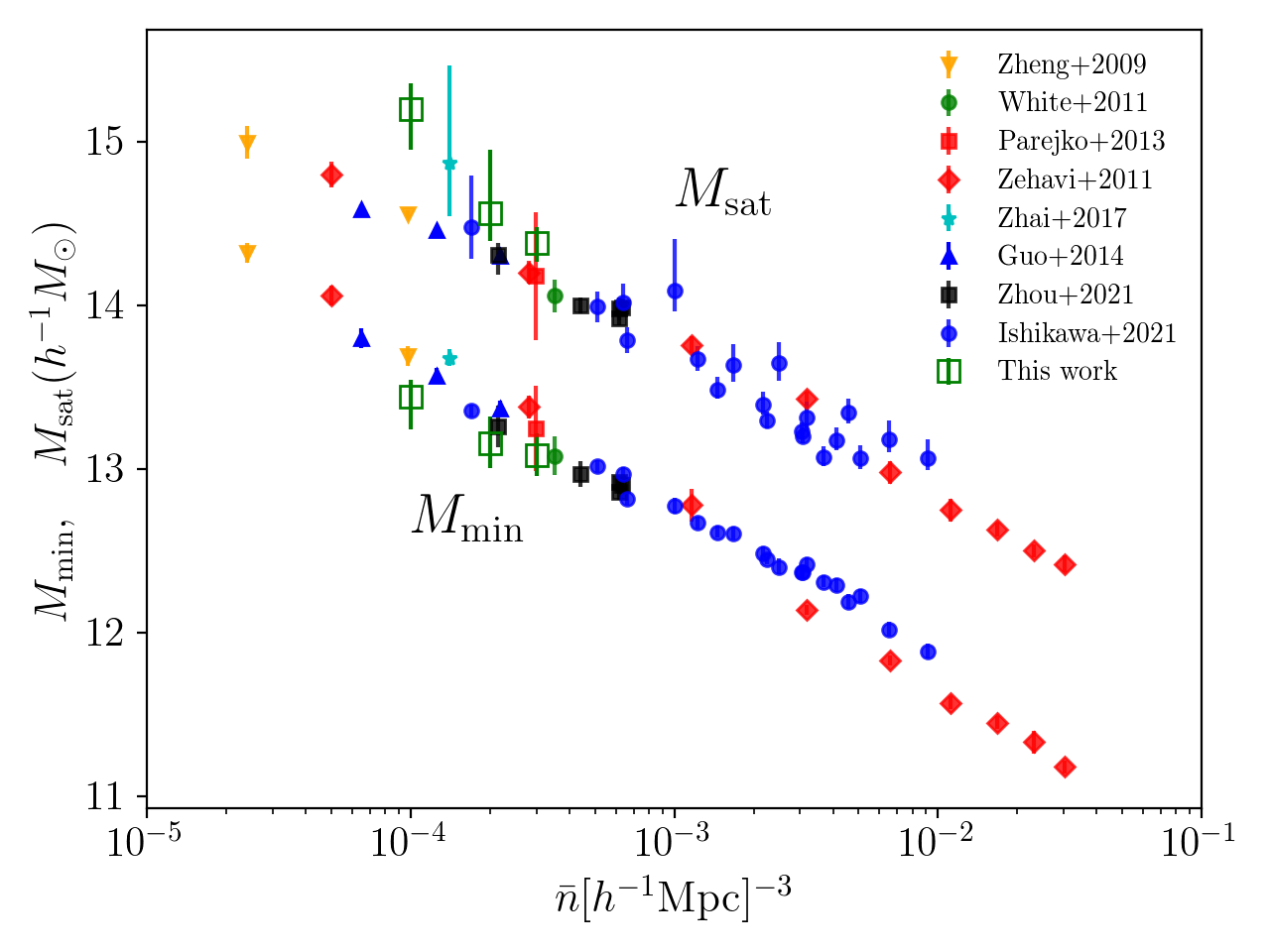}
\caption{HOD parameters $M_{\text{min}}$ and $M_{\text{sat}}$ as a function of galaxy number density from our luminosity dependent analysis, compared with other investigations including SDSS (\citealt{Zheng_2009, CMASS_Martin, Zehavi_2011, Parejko_LOWZ, Guo_2014, Zhai_2017}), DESI (\citealt{Zhou_2021}) and HSC (\citealt{Ishikawa_2021}). The overall agreement of these constraint approves the correlation between halo mass and galaxy number density.}
\label{fig:Mmin_Msat}
\end{center}
\end{figure}

Amongst the HOD parameters extracted from the clustering analysis, $M_{\text{min}}$ evaluates the halo mass scale at which the expected central occupancy reaches 50\%, conditional on the $f_{\text{max}}$ parameter in our model, while $M_{\text{sat}}$ represents a typical halo that can host one satellite galaxy. We use the result from our luminosity dependent (Section~\ref{sec:luminosity}) analysis and project the measurement as a function of galaxy number density in Figure~\ref{fig:Mmin_Msat}. We compare this result with other investigations using similar HOD-based method, including samples from SDSS (\citealt{Zheng_2009, CMASS_Martin, Zehavi_2011, Parejko_LOWZ, Guo_2014, Zhai_2017}), DESI (\citealt{Zhou_2021}) and HSC (\citealt{Ishikawa_2021}). Note that these analyses use galaxy samples at different redshifts, and different HOD models. However, these results reveal a clear dependency on luminosity. Our brightest subsample N1 shows a somewhat higher value of $M_{\text{sat}}$, which is likely due to the low satellite fraction of the brightest galaxies. In addition, our result has slightly larger uncertainties for both $M_{\text{min}}$ and $M_{\text{sat}}$, since our measurement is the only one that can marginalize over cosmological parameters.

In our analysis, galaxy subsamples are selected by their rank-ordering in thin redshift slices until they reach the threshold of number densities. When this selection starts from two ends, the high number density threshold unavoidably returns subsamples with overlap. This effect is quite minor for the N1 subsamples. In Figure \ref{fig:nz}, only a small amount of overlap exists for galaxies with $z>0.61$. However, this overlap is significant for N2 subsamples. At $z>0.61$, all galaxies can be selected from both extremes and can make the comparison moot. It is less significant for lower redshift. A zero-overlap result requires the number density of the original galaxy catalog to be at least twice the threshold which is only satisfied at $z\sim0.5$ for BOSS. Therefore, The comparison with N2 subsample therefore needs undertaken with care. The overlap in samples may lead to clustering measurements that are less distinct and because the clusering emasurements from the subsamples are correlated. 

\section{Conclusions} \label{sec:conclusion}

Galaxy clustering on intermediate scales is determined by both cosmology and galaxy formation processes. Accurate modeling of this signal can be used to better constrain both cosmological and galaxy bias. After marginalizing over the galaxy bias parameters, we find that the clustering measurement at small to intermediate scales provides a strong constraint on the growth rate of structure. In Z22, we applied this method to the BOSS galaxies and measured the growth rate parameterized by $f\sigma_{8}$ over a redshft range $0.18<z<0.62$ with an uncertainty of 5.5\% to 7.8\%, better than the measurement from large, linear scales. However these results are lower than the prediction assuming a Planck cosmology with a tension of up to $3.4\sigma$. In order to investigate the robustness of this small scale analysis, as well as explore any dependence on the galaxy properties or observational systematics, we focus on the BOSS galaxies at $0.48<z<0.62$ and perform an extended analysis considering subsamples of galaxies with different properties. 

For this purpose, we have constructed emulators for the galaxy correlation function based on the HOD formalism at three number densities lower than the BOSS density used in Z22, and have used them to fit the corresponding galaxy samples. Given the Granada VAC for BOSS galaxies, we have access to galaxy properties including magnitude, color, age, stellar mass, and sSFR. We define galaxy samples using these parameters with different thresholds of number density, resulting  in samples with different average galaxy properties. The measurement of the galaxy correlation function reveals the well-known dependence of clustering: Brighter, redder, older, more massive and lower sSFR galaxies are more strongly clustered than fainter, bluer, younger, less massive and higher sSFR galaxies. This result implies intrinsic correlations among these galaxy properties but some scatter is possible at various scales. 

We fit to the measured clustering from these subsamples using emulator predictions and a Bayesian analysis, and measure both cosmological parameters and HOD / galaxy bias parameters. For $f\sigma_{8}$ the measurement is consistent, regardless of how we split galaxies by their properties. Compared with Z22, the most significant deviation is less than $2\sigma$ evaluated using Eq.~(\ref{eq:tension}), and they are all consistently lower than Planck. If this tension between the measured growth rate and that predicted by Planck comes from a problem with the $\Lambda$CDM model, then this has to be independent of the other cosmological parameters, which seems unlikely: see similar discussions in \cite{Amon_2022} and \cite{Chapman_2023} where the authors investigate possible explanations based on baryonic processes, or a deviation in the non-linear growth rate of structure leading to changes on small-scales.

We then interpret the clustering measurement of different galaxy samples within the HOD model. This shows that, as expected, the clustering dependence can be largely explained by the halo mass. Galaxies residing in more massive halos have an overall higher clustering amplitude, such as brighter, redder, older, more massive and quenched galaxies. In addition to halo mass, we also find some secondary dependencies. For instance, brighter galaxies can have higher velocity bias compared to fainter galaxies. But this tendency is not observed in samples selected by color, age or stellar mass. This indicates scatter of the parameters describing galaxy formation and distribution. 

To complete the analysis, we repeat our fits for galaxy samples selected by additional Stellar-Population parameters, which are less intuitive but also correlated with physical parameters. Using look-back formation time, we find that the subsamples give different clustering amplitudes with a similar but weaker influence. Metallicity, on the other hand, doesn't give distinct clustering measurements and the selection is close to a random downsampling due to reduced number density. The constraints on $f\sigma_{8}$ are quite consistent with previous results.

Our analysis within the HOD framework shows some scatter for the empirical parameters and galaxy properties beyond that expected from statistical scatter. It is also likely that an improvement in the HOD modeling is necessary to better describe the data and model. Research along this line has been visited in literature, see for instance the correlation between galaxy size and dark matter halos (\citealt{Hearin_2017}), color distribution in the HOD model (\citealt{Hearin_2013,Campbell_2015}) and so on. Incorporating these effects into future models will introduce additional degrees of freedom, but also allows more comprehensive exploration for the relationship between galaxies and dark matter halos. This work will need elaborate considerations when both galaxy bias and cosmology are allowed to vary, but can be possible given our high resolution simulations and we will leave it in future investigations.

% End of mnras_template.tex

\section*{Acknowledgements}

WP acknowledges the support of the Natural Sciences and Engineering Research Council of Canada (NSERC), [funding reference number RGPIN-2019-03908]. HG acknowledges the support of National Science Foundation of China (Nos. 11922305, 11833005) and the science research grants from the China Manned Space Project with NO. CMS-CSST-2021-A02.

Research at Perimeter Institute is supported in part by the Government of Canada through the Department of Innovation, Science and Economic Development Canada and by the Province of Ontario through the Ministry of Colleges and Universities.

This research was enabled in part by support provided by Compute Ontario (computeontario.ca) and the Digital Research Alliance of Canada (alliancecan.ca).

\rm{Software:} Python,
Matplotlib \citep{matplotlib},
NumPy \citep{numpy},
SciPy \citep{scipy},
George \citep{george_2014},
Corrunc \citep{Sinha_2020},
MultiNest (\citealt{Feroz_2009, Buchner_2014}).

\section*{Data Availability}

The code and data are available by request to the authors.

\bibliographystyle{mnras}
\bibliography{emu_gc_bib,software}

\appendix

\section{Dependence on covariance matrix} \label{appsec:covariance}

We use the Gaussian kernel method to smooth the data covariance matrix throughout our work. In order to test if the smoothing method can have an impact on the cosmological inference, we repeat the likelihood analysis for the brightest sample with N1 number density (Section~\ref{sec:luminosity}) considering different methods to estimate the covariance matrix. We consider the unsmoothed covariance matrix multiplied by the \cite{Hartlap_2007} factor (Hartlap Correction), the factor suggested by \cite{Percival-2022} (Percival Correction), or a revised measurement of the Jackknife split following \citet{Mohammad_2022} (Corrected Jackknife). Figure~\ref{fig:covariance_matrix_test} shows the constraint on the cosmological parameters with these different methods. We can see that the offset due to the correction on the covariance matrix is quite minor.

\begin{figure}
\begin{center}
\includegraphics[width=8cm]{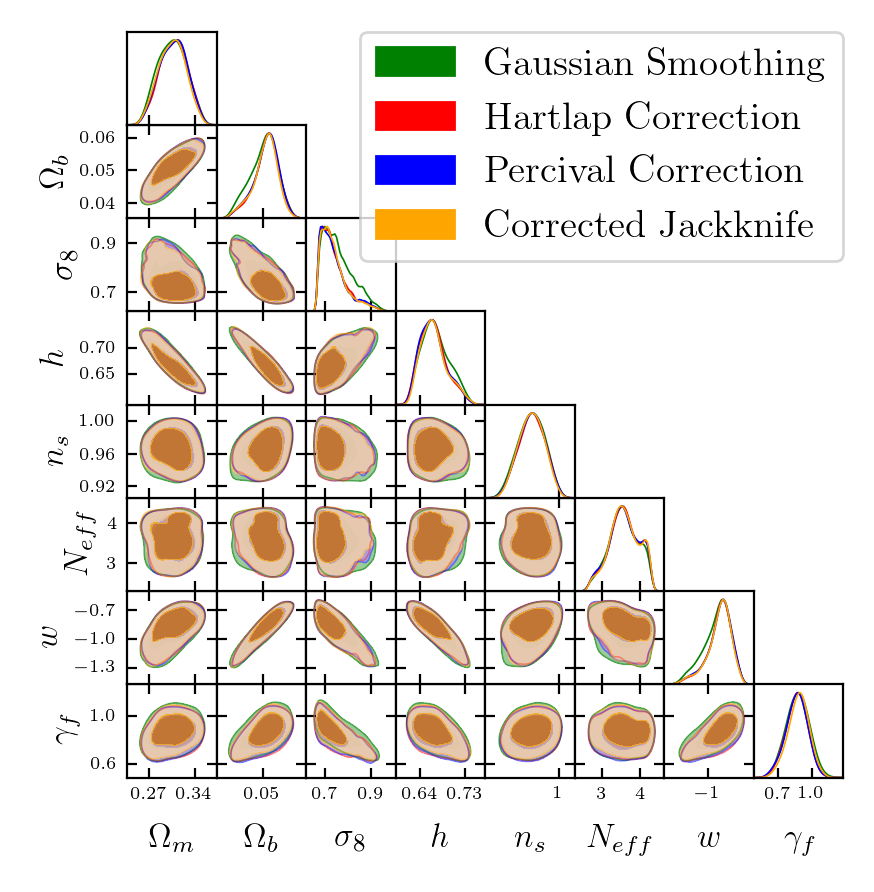}
\caption{Test for different algorithms for determining the covariance matrix. We use the galaxies selected by i-magnitude with N1 number density, i.e. the averagely brightest galaxy sample. The result shows the constraint on the cosmological parameters with negligible impact from the covariance matrix.}
\label{fig:covariance_matrix_test}
\end{center}
\end{figure}

\section{Measurement of correlation function from Granada VACs}\label{appsec:CF}

The Granada VACs employ various templates for priors on star formation-time scenarios, dust model and initial mass function, which can affect the derived properties of individual galaxies. We do not find qualitative changes in the resulting clustering and the model fits to it, if we consider different priors. For example, Figure~\ref{fig:BOSS_CF_color} shows the measurement for galaxies selected by color. Red galaxies are more clustered than blue galaxies, consistent with earlier analysis at lower redshift (\citealt{Zehavi_2011}), for all priors. The measurements for other galaxy properties give a similar result, i.e. the clustering measurement is insensitive to the choice of template.

\begin{figure*}
\begin{center}
\includegraphics[width=17cm]{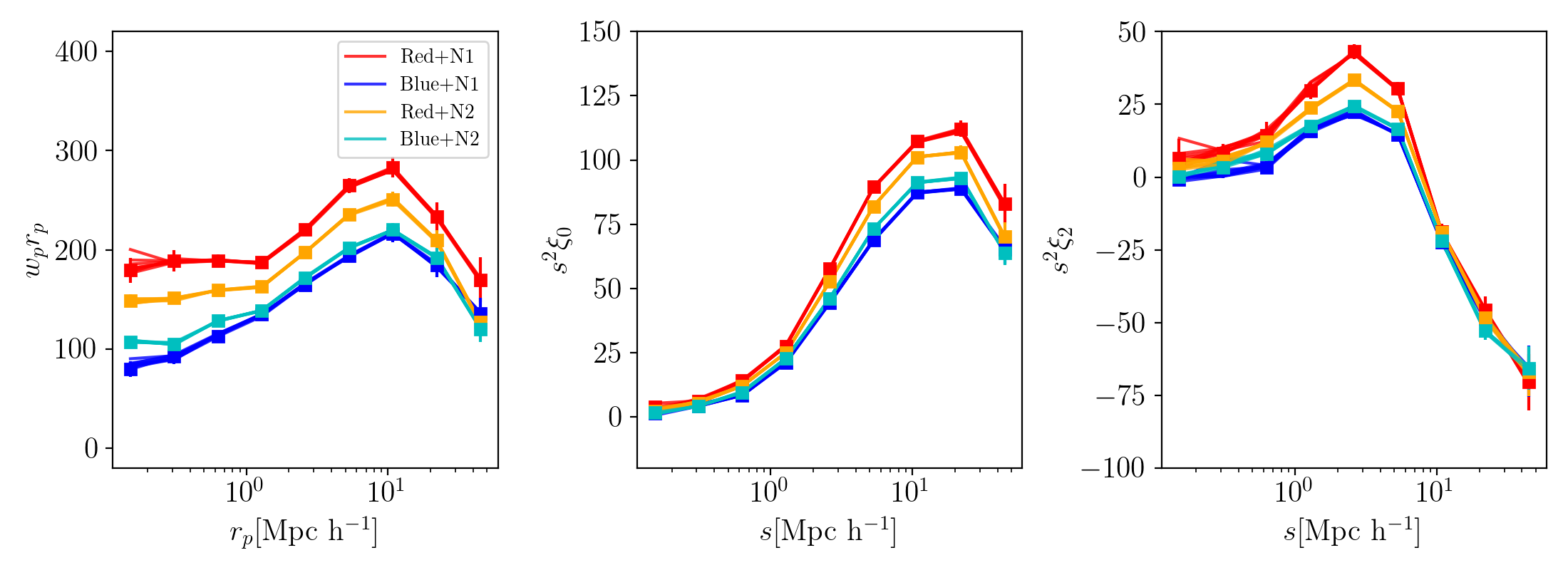}
\caption{Measurement of galaxy correlation function for galaxy samples selected by color and number density. Due to the flexibility of the Granada VAC, we present results from all eight versions of the catalog, corresponding to the choices of star formation-time scenarios, dust and initial mass functions. These parameters can impact the properties of individual galaxies, however their relative positions in the catalog are not significantly affected and thus can yield consistent measurements of correlation function. We also observe this consistency when galaxies are selected by other parameters as analyzed in the text. The errorbars in the figure correspond to the Jackknife-resampling uncertainty from one of the catalogs.}
\label{fig:BOSS_CF_color}
\end{center}
\end{figure*}

\section{Constraint on the model parameters}

We summarize the constraint on a subset of the model parameters in Table~\ref{tab:constraint}. Explanations can be found in the context.

\begin{landscape}
\begin{table}  
\caption{Measurement of a subset of the model parameters, including velocity bias for centrals ($\eta_{\mathrm{vc}}$) and satellites ($\eta_{\mathrm{vs}}$), concentration parameter for satellite distribution ($\eta_{\text{con}}$), velocity scaling parameter $\gamma_{f}$, amplitude of the assembly bias parameter $f_{\text{env}}$, asymptotic value of central occupancy $f_{\text{max}}$, satellite fraction $f_{\text{sat}}$ and $\chi^{2}$ for each fit.} \label{tab:constraint}
\begin{tabular}{rcccccccccccc}
\cline{1-13}
  Parameter      & $\log[M_{\mathrm{sat}}(h^{-1}M_{\odot})]$ & $\alpha$ & $\log[M_{\mathrm{cut}}(h^{-1}M_{\odot})]$  &  $\sigma_{\log M}$  &  $\eta_{\mathrm{vc}}$  &  $\eta_{\mathrm{vs}}$  &  $\eta_{\mathrm{con}}$  &  $\gamma_{f}$  &  $f_{\mathrm{env}}$  &  $f_{\mathrm{max}}$  & $f_{\mathrm{sat}}$ & $\chi_{2}$\\
\cline{1-13}
N1 & $15.2\pm0.2$ & $0.74\pm0.18$ & $11.8\pm1.01$ & $0.48\pm0.15$ & $0.32\pm0.06$ & $1.26\pm0.2$ & $0.65\pm0.23$ & $0.87\pm0.1$ & $0.08\pm0.08$ & $0.78\pm0.21$ & $0.06\pm0.01$ & 20.6 \\
N2 & $14.56\pm0.28$ & $1.1\pm0.33$ & $11.86\pm1.16$ & $0.4\pm0.16$ & $0.26\pm0.06$ & $0.92\pm0.12$ & $0.73\pm0.25$ & $0.88\pm0.11$ & $-0.01\pm0.09$ & $0.63\pm0.17$ & $0.08\pm0.01$ & 11.4 \\ 
N3 & $14.38\pm0.11$ & $1.12\pm0.15$ & $12.04\pm1.24$ & $0.39\pm0.16$ & $0.24\pm0.06$ & $0.87\pm0.08$ & $0.65\pm0.18$ & $0.86\pm0.09$ & $-0.04\pm0.06$ & $0.75\pm0.14$ & $0.09\pm0.01$ & 9.3 \\
\cline{1-13}
Blue+N1 & $14.67\pm0.16$ & $1.52\pm0.26$ & $11.73\pm1.11$ & $0.37\pm0.18$ & $0.31\pm0.08$ & $0.76\pm0.19$ & $0.94\pm0.59$ & $0.82\pm0.11$ & $-0.0\pm0.1$ & $0.18\pm0.04$ & $0.06\pm0.01$ & 6.2 \\
Blue+N2 & $14.36\pm0.1$ & $1.29\pm0.22$ & $12.51\pm1.38$ & $0.41\pm0.14$ & $0.29\pm0.09$ & $0.79\pm0.14$ & $0.78\pm0.4$ & $0.91\pm0.09$ & $-0.06\pm0.06$ & $0.35\pm0.06$ & $0.08\pm0.01$ & 12.9 \\
Red+N2 & $14.38\pm0.11$ & $1.32\pm0.2$ & $11.7\pm1.3$ & $0.37\pm0.16$ & $0.31\pm0.07$ & $0.88\pm0.1$ & $0.79\pm0.29$ & $0.83\pm0.09$ & $-0.07\pm0.08$ & $0.5\pm0.09$ & $0.09\pm0.01$ & 7.3 \\
Red+N1 & $14.53\pm0.12$ & $1.36\pm0.19$ & $11.64\pm1.18$ & $0.34\pm0.19$ & $0.29\pm0.08$ & $0.98\pm0.13$ & $0.71\pm0.39$ & $0.83\pm0.11$ & $-0.08\pm0.13$ & $0.3\pm0.08$ & $0.1\pm0.02$ & 8.8  \\
\cline{1-13}
Young+N1 & $14.73\pm0.21$ & $1.43\pm0.29$ & $11.44\pm1.17$ & $0.39\pm0.16$ & $0.3\pm0.1$ & $0.74\pm0.17$ & $0.73\pm0.56$ & $0.89\pm0.1$ & $0.12\pm0.1$ & $0.13\pm0.02$ & $0.06\pm0.01$ & 12.6\\
Young+N2 & $14.46\pm0.16$ & $1.23\pm0.28$ & $12.22\pm1.21$ & $0.46\pm0.13$ & $0.34\pm0.07$ & $0.9\pm0.13$ & $0.65\pm0.32$ & $0.87\pm0.11$ & $-0.06\pm0.06$ & $0.35\pm0.07$ & $0.08\pm0.01$ & 8.9\\ 
Old+N2 & $14.41\pm0.12$ & $1.23\pm0.23$ & $11.97\pm1.32$ & $0.36\pm0.16$ & $0.28\pm0.06$ & $0.84\pm0.09$ & $0.82\pm0.29$ & $0.85\pm0.09$ & $-0.05\pm0.08$ & $0.6\pm0.11$ & $0.09\pm0.01$ & 11.7 \\
Old+N1 & $14.61\pm0.15$ & $1.34\pm0.21$ & $11.66\pm1.23$ & $0.41\pm0.17$ & $0.26\pm0.08$ & $0.91\pm0.13$ & $0.79\pm0.35$ & $0.82\pm0.11$ & $-0.04\pm0.12$ & $0.49\pm0.13$ & $0.08\pm0.01$ & 8.8 \\
\cline{1-13}
Less Massive+N1 & $14.95\pm0.28$ & $1.13\pm0.29$ & $11.92\pm0.96$ & $0.48\pm0.13$ & $0.42\pm0.07$ & $0.85\pm0.29$ & $1.06\pm0.6$ & $0.89\pm0.1$ & $0.04\pm0.08$ & $0.12\pm0.02$ & $0.06\pm0.01$ & 16.1\\
Less Massive+N2 & $14.41\pm0.1$ & $1.37\pm0.23$ & $11.68\pm1.31$ & $0.44\pm0.13$ & $0.31\pm0.08$ & $0.81\pm0.13$ & $1.03\pm0.47$ & $0.88\pm0.09$ & $-0.05\pm0.06$ & $0.28\pm0.05$ & $0.08\pm0.01$ & 8.9\\
More Massive+N2 & $14.42\pm0.14$ & $1.24\pm0.25$ & $12.04\pm1.34$ & $0.37\pm0.17$ & $0.25\pm0.07$ & $0.86\pm0.09$ & $0.81\pm0.3$ & $0.88\pm0.11$ & $-0.05\pm0.07$ & $0.62\pm0.13$ & $0.09\pm0.01$ & 8.0\\
More Massive+N1 & $14.98\pm0.22$ & $0.92\pm0.25$ & $11.68\pm1.12$ & $0.41\pm0.14$ & $0.29\pm0.06$ & $1.01\pm0.16$ & $0.47\pm0.17$ & $0.78\pm0.1$ & $0.03\pm0.1$ & $0.79\pm0.17$ & $0.07\pm0.01$  & 20.7\\
\cline{1-13}
Quenched+N1 & $14.77\pm0.18$ & $1.19\pm0.2$ & $11.27\pm0.96$ & $0.34\pm0.18$ & $0.2\pm0.09$ & $0.92\pm0.12$ & $0.34\pm0.14$ & $0.84\pm0.11$ & $0.07\pm0.09$ & $0.44\pm0.13$ & $0.08\pm0.02$ & 17.6\\
Quenched+N2 & $14.36\pm0.1$ & $1.34\pm0.22$ & $11.64\pm1.39$ & $0.39\pm0.15$ & $0.3\pm0.07$ & $0.82\pm0.1$ & $0.99\pm0.33$ & $0.87\pm0.1$ & $-0.05\pm0.08$ & $0.54\pm0.11$ & $0.09\pm0.01$ & 8.2\\
Star Forming+N2 & $14.64\pm0.25$ & $0.99\pm0.28$ & $12.11\pm1.22$ & $0.42\pm0.15$ & $0.31\pm0.07$ & $0.9\pm0.12$ & $0.68\pm0.35$ & $0.82\pm0.1$ & $-0.06\pm0.07$ & $0.39\pm0.09$ & $0.09\pm0.01$ & 7.5 \\
Star Forming+N1 & $14.75\pm0.18$ & $1.34\pm0.22$ & $11.33\pm1.24$ & $0.42\pm0.15$ & $0.34\pm0.08$ & $0.94\pm0.15$ & $0.68\pm0.39$ & $0.83\pm0.09$ & $0.01\pm0.09$ & $0.17\pm0.03$ & $0.07\pm0.01$ & 18.1\\
\cline{1-13}
\end{tabular}
\end{table}

\end{landscape}

%\bsp	% typesetting comment
\label{lastpage}
\end{document}